\documentclass[onecolumn]{mnras}

\usepackage{graphicx}
\usepackage{longtable}
\usepackage{newtxtext,newtxmath}
\usepackage[T1]{fontenc}
\usepackage{amsmath}	% Advanced maths commands

\usepackage{amssymb}	% Extra maths symbols
\usepackage{mathtools}

\usepackage{natbib}
\usepackage{pdflscape}

\title[Cataclysmic Variables Photometric Periods from TESS] {Cataclysmic Variables Photometric Periods from TESS} 
\author[S. O. Kepler et al.] { S. O. Kepler$^{1}$, Alejandra D. Romero$^{1}$, L. L. Amorim$^{1}$, Marcos P. Diaz$^{2}$ , Laura C. Sultan$^1$\\
$^{1}$Instituto de F\'{\i}sica, Universidade Federal do Rio Grande do Sul, 91501-970 Porto Alegre, RS, Brazil\\
$^{2}$Universidade de S\~ao Paulo, Instituto de Astronomia, Geof\'isica e Ci\^encias Atmosf\'ericas, Rua do Mat\~ao, 1226, S\~ao Paulo, SP 05508-900, Brazil }

\date{Accepted XXX. Received XXX; in original form XXX}

\begin{document}

\maketitle

%\shorttitle{A TESS view of CVs}
%\shortauthors{Kepler et al.}

%% Mark off the abstract in the ``abstract'' environment. 
\begin{abstract}

We present a sample of coherent and stable photometric period determinations for cataclysmic variables, based on TESS photometry through sector 102. We analyzed a total of 1554 cataclysmic variable stars and detected periodic variations in 1362 objects, including 286 eclipsing or ellipsoidal-variation systems, 63 polars, and 135 intermediate polars. In particular, we present the first determination of the optical variability period for 564 cataclysmic variables. Due to the 21$\times$ 21" pixel size of TESS, we tested whether the variability was coming from the cataclysmic variable and not from a nearby star.
For the intermediate polars, we detected spin periods in addition to orbital periods for 83 systems. 
We detect a clear period gap between $\sim$2 and $\sim$ 3~h in the eclipsing sample, consistent with previous work. The gap remains for the complete sample of photometric variability periods. There is no apparent gap in the orbital period distribution for intermediate polars. Finally, the median of the photometric period distribution of our complete sample of cataclysmic variables is 3.681~h. Comparing this to the distribution of rotation periods for likely single white dwarfs, we find a similar range, with a median photometric period of 6.803~h. 
\end{abstract}


\section{Introduction} 
\label{sec:intro}

Cataclysmic variables (CVs) are semidetached binary systems in which a compact star, the primary, is a white dwarf that is accreting material and angular momentum from the secondary, usually a low-mass main-sequence star filling its Roche lobe \citep[see][for a review]{1995cvs..book.....W}, although it can occasionally be a giant. The mass transfer from the donor occurs primarily through the formation of an accretion disk around the white dwarf and a hotspot where the gas stream impacts the disk. For magnetic CVs, the accretion disk may be entirely suppressed or truncated at the white dwarf's magnetospheric radius, forming polar and intermediate polar (IP) CVs, respectively. 

Most systems show orbital periods of a few hours, with a few exceptions for binaries with periods of a few days. These systems exhibit diverse accretion modes, leading to a wide range of photometric variability timescales  --- from seconds to minutes, up to months or even years --- and thus serve as ideal laboratories for studying binary-star accretion. CVs are also among the most common end products of binary star evolution, granting them a prominent position in the field of binary star evolution studies.
The luminosity of a CV can vary significantly, and systems can be in low-luminosity (quiescence), high-luminosity (outburst), or intermediate states when active but not at peak luminosity. Flickering, originating from the shocked region where the mass-transfer stream impacts the outer rim of the disk, the hotspot, or the inner regions of the disk, is common.

 Since white dwarf stars are the most common end products of stellar evolution, and accretion disks are the most common universal structures resulting from angular-momentum-driven mass transfer, understanding the accretion process in CVs is a step toward a comprehensive understanding of accretion in other systems.

The evolution of interacting binary systems, including cataclysmic variables, is strongly dominated by angular momentum loss \citep{1983ApJ...275..713R, 2004ApJ...601.1058I, 2011ApJS..194...28K}. For orbital periods longer than $\sim 3$~h, the angular momentum loss is dominated by efficient magnetic braking. When the orbital period is around 3\,h, the donor star in the CV is fully convective, and magnetic braking is significantly reduced. The donor star, slightly bloated from prior mass loss, shrinks to its thermal equilibrium radius, detaches from the Roche lobe, and halts mass transfer. The CV is now a detached WD+MS binary.  For periods below $\sim$3~h, gravitational radiation is the primary mechanism for angular momentum loss, which reduces the orbit sufficiently for the donor to fill its Roche lobe and for the system to resume accretion as a CV when the orbital period is $\sim$~2~h. This leads to a period gap from  2.15 to 3.18\,h, where the number of CVs is drastically reduced \citep[e.g.][]{1983ApJ...275..713R,2010A&A...513L...7S, 2016MNRAS.457.3867Z, 2023MNRAS.524.4867I}.  The orbital period continues to shorten until it reaches the period minimum of $\sim$~80~min \citep{1981ApJ...248L..27P,1982ApJ...254..616R, 2009MNRAS.397.2170G,2020MNRAS.494.3799P}. There are problems with the observed mass transfer rates in systems below the 2--3 hour period gap, which are often several times higher than predicted by general relativity alone, suggesting the need for additional angular momentum loss mechanisms \cite[e.g.][]{Liu_Li_2018,Zhou26}.

Some cataclysmic variables exhibit several photometric variability periods, including the orbital period, the spin period --- if synchronized ---, several superhump periods, and their harmonics.
Superhumps are brightness variations that occur at a period either slightly longer (positive) or slightly shorter (negative) than the orbital period, caused by 
variations in the brightness of an eccentric accretion disk that undergoes prograde apsidal precession or by an inclination or tilt of the accretion disk with respect to the orbital plane, together with retrograde nodal disk precession. The beat period between the orbit and the superhumps is the disk precession period \cite[e.g.][]{Bruch23a}. 
The orbital period distribution of CVs can provide important information on the physical properties of these systems and their evolution. Since the discovery of the first cataclysmic variable, U Geminorum, by \cite{1856MNRAS..16...56H}, several authors have worked to determine the orbital period distribution for CVs. \citet{1995xrbi.nasa..578R} presented a compilation of CVs with known periods, including X-ray binaries and related objects, with data gathered from previous works \citep[][for instance]{1984A&AS...57..385R, 1987MPARp.285.....R, 1990A&AS...85.1179R}. The following updates were published in \citet[][update RKcat7.24, 2016]{2003A&A...404..301R}, where the authors presented 472 cataclysmic variables with known or suspected orbital periods 

The catalog by \citet{2001PASP..113..764D} lists 1034 CVs as of December 2000.
Period information was available for approximately one-third of the objects and was taken from \citet{1998A&AS..129...83R}. 

\citet{Schaefer22} searched for orbital periods of known nova systems by identifying significant, coherent, and stable optical photometric modulation in data from TESS, Kepler, AAVSO, SMARTS, OGLE, ASAS, and ZTF. He presented a list of 156 orbital periods, including 49 new periods. \citet{Schaefer24} discusses the failure of predictions from magnetic braking models to explain the observed orbital period distribution; \cite{Jorquera25} proposed saturated and disrupted models, and \cite{Barraza26} introduced updated models by showing that only moderate changes in angular momentum loss at the fully convective boundary are required to explain the orbital period distribution. \cite{Shi26} proposes that magnetic braking and common envelope physics must be considered together, while \cite{Dodon26} proposes that including irradiation effects on the magnetic braking prescription can explain the observed period gap.

 \citet{2023AJ....165..163C} used a sample of 1587 known CVs to study the spatial distribution and luminosity function of CVs in our Galaxy. They used astrometric and photometric data from Gaia DR3 to determine distances, while orbital periods were taken from \citet{2003A&A...404..301R} and \citet{2001PASP..113..764D}. In their sample, 767 objects had measured orbital periods.

\citet{2023MNRAS.524.4867I} and \citet{2025MNRAS.536.1057I} presented a catalog of 507 CVs using optical spectra from SDSS~I to V, combined with Gaia astrometry. They also searched the Catalina, ZTF, and TESS catalogs for photometric data to determine the orbital periods.  They reported 326 periods, including 59 new ones. 

In this work, we study the photometric periods of CVs using data from the TESS space telescope \citep{Ricker15}.  TESS observes the sky in $24\deg\times 96\deg$ fields called sectors, with a pixel angular size of $21\times 21$ arcsec.
Each sector comprises two orbits of the satellite around the Earth, lasting about 27 days on average. The TESS passband covers from 6000 to 10000~\AA, centred on the Cousins I band.
TESS collects 2-minute postage-stamp observations for up to 20,000 selected stars per sector
due to the limited Ka-band downlink bandwidth, restricted perigee contact windows, and onboard solid-state storage capacity. The SPOC pipeline \cite{2020RNAAS...4..201C} processed these postage stamps, calibrating the pixels, extracting photometry and centroids, removing instrumental signatures, and producing 2-min light curves. Because of the large pixel size, multiple stars can contribute to the same pixel and create blended signals \cite{Guerrero21}.  There are gaps in the light curves within sectors and between observing campaigns.  Spacecraft jitter, scattered light from Earth or the Moon, momentum dumps, and other instrumental effects can introduce noise in the data. In addition, the 2-minute target list is not a random sample because targets are prioritised using brightness, stellar type, scientific programmes, and technical constraints \cite{Fausnaugh21}, and therefore, the selection cannot be treated as a complete or statistically representative population \cite{Kunimoto23}.

We visually inspected each 2-minute light curve and Fourier spectrum to identify significant periods and to test whether they truly originate from the system rather than from a nearby star. We compiled photometric orbital periods for 789 CVs previously classified as such in the literature. As we study only photometry, not spectra, we do not attempt to classify the detected periodicities; instead, we list the coherent periodicities across several TESS sectors. 
Unlike orbital and spin periods, other photometric periods are less stable because they can be affected by mass transfer or magnetic fields.
Sample selection and data reduction are presented in Section~\ref{section2}. We analyse the main photometric period distribution for the sample of CVs in general and, in particular, for eclipsing and magnetic systems in Section~\ref{section3}. We also compare our results with those in the literature and with the distribution of rotation periods for single, non-interacting white dwarfs. We summarise our results in Section~\ref{summary}.

\section{Sample and data analysis}
\label{section2}

The sample of CVs presented in this work was constructed as follows. We searched for known CV systems in the TESS data using catalogues from the literature, mainly those presented by \citet{2001PASP..113..764D,Schaefer22, 2023MNRAS.524.4867I,  2025MNRAS.536.1057I, 2023AJ....165..163C,Bruch25, Green25} and \citet{Dag26}.  

Using these catalogues as a starting point, we searched for 120~s and 20~s light curves for all identified CVs from The Mikulski Archive for Space Telescopes, hosted by the Space Telescope Science Institute (\textsc{STScI}) \citep{https://doi.org/10.17909/t9-st5g-3177, https://doi.org/10.17909/t9-tcn7-7g94, https://doi.org/10.17909/t9-nmc8-f686, https://doi.org/10.17909/t9-yk4w-zc73}\footnote{http://archive.stsci.edu/} in \textsc{FITS} format. This process yielded a sample of 1554 objects observed by \textrm{TESS} across sectors 1---102.

 We use the \textsc{Lightkurve} package \citep{lk18} to download the photometric data observed by \textrm{TESS}. We selected data taken at 120~s and 20~s cadences, when available, and processed them with the \textsc{SPOC} pipeline \citep{2020RNAAS...4..201C}. 
 The \textrm{TESS} telescope observes each sector for $27$~days, interrupting observations every $\sim 13$~days -- or less -- to downlink the data to Earth. 
 %Due to these gaps in the light curve, we searched for stable variability only for periods of up to 13 days. 
After a 5$\sigma$ clip, we normalised the per-sector light curves individually, combined the light curves from all sectors where the star was observed, and then computed a complete Fourier transform (FT) suitable for unevenly spaced data to detect coherent variability signals. The FT advantage over other periodogram methods lies in its significance levels and alias structure, which are well-known and standard \citep[e.g.,][]{Vio10}. We tested periods from 240~s (the Nyquist for the 120~s datasets, 40~s for the 20~s fast datasets) to around 13~d, set by the length of the uninterrupted part of the TESS light curves, except when the folded light curve showed eclipses at longer periods. When available, we also analysed the 20~s datasets, which are important for periods shorter than $\sim 400$~s.
The detection limit is computed using the false-alarm probability (FAP=1/1000), calculated by reshuffling the data 1000 times, maintaining the same time base, computing the FT, and analysing all peaks above the fap. As shown in \cite{1993BaltA...2..515K},
the false-alarm probability calculated by reshuffling the data points does not assume that the noise is randomly distributed.
Statistically, a peak with an amplitude above the FAP in the FT has less than a one-in-a-thousand chance of being due to noise. It is important to note that false-alarm probabilities are only reliable for coherent variations in light curves, whereas CVs have outbursts and flickering that can dominate over a weak photometric period.
For TESS data, the value of the FAP=1/1000 is on the order of 4$\left<A\right>$ for 120\,s--cadence and 5$\left<A\right>$ for 20\,s--cadence data \citep[see][for instance]{2014ApJ...794...39B, 2016A&A...585A..22Z}, where $\left<A\right>$ is the standard deviation of the background noise to the average background signal. For each object in the sample, we use the TESS-Localize software
\footnote{https://github.com/Higgins00/TESS-Localize}
\citep{2023AJ....165..141H} to ensure that the detected period originates from the system of interest. Note that previous works using TESS data \citep{Bruch22,Bruch23a,Bruch23b,Bruch24a,Bruch24b,Bruch25,Hernandez25,Dag26} did not test whether the periodicity they detected actually came from the CV.
%and in at least 7 cases (see Section~\ref{swrong}), they reported the period of a nearby eclipsing binary as that of the CV itself.

\begin{figure}
    \includegraphics[width=0.90\textwidth]{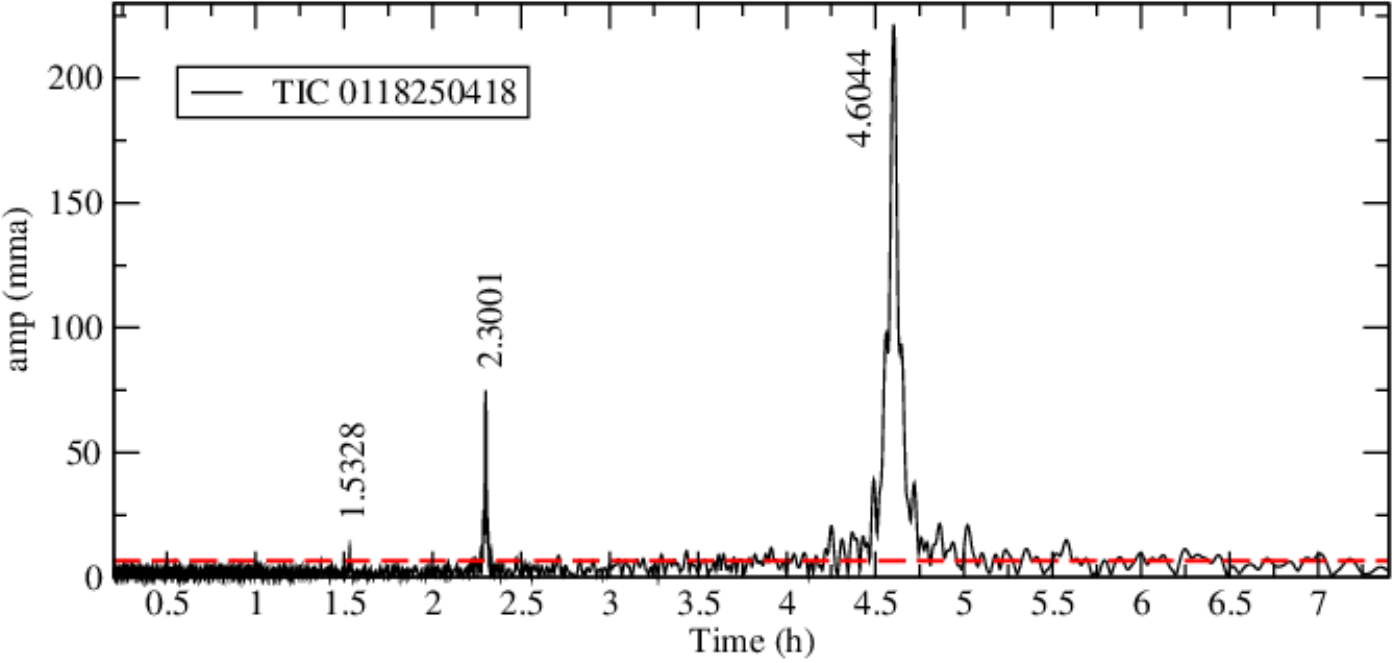}
            \caption{Fourier Transform for TIC~0118250418, considering the two sectors observed. The periods are in hours, and the peaks are indicated in the figure. The horizontal red line corresponds to the false-alarm probability FAP=1/1000 detection limit. }
    \label{0118250418}
\end{figure}

For each object, we determine the photometric periods by carefully analysing the FT and the light curve. \citet{Schaefer22}
proposes that any periodicity in the range 0.04--10~d that is coherent, stable, and signiﬁcant must be tied to a very good clocking mechanism in the binary, which can only be the orbital period. We do not assume the detected periods are orbital because TESS is a small telescope with large pixels, which can cause strong contamination in some fields, resulting in significant noise on timescales of several days.
Several works in the literature assume that the largest-amplitude peak detected is the orbital period of CVs, except for a few cases \citep[e.g.,][]{Bruch22,Bruch23a,Bruch23b,Bruch24a,Bruch24b,Bruch25,Hernandez25,Dag26}. As an example, the FT for TIC~0118250418 is shown in Figure~\ref{0118250418}, where we identify three peaks above the detection limit, indicated by the horizontal red line. For this object, we assume the main period is 4.6044~h, which corresponds to the dominant period. The first and second harmonics are also present.   

\begin{figure}
        \includegraphics[width=0.85\textwidth]{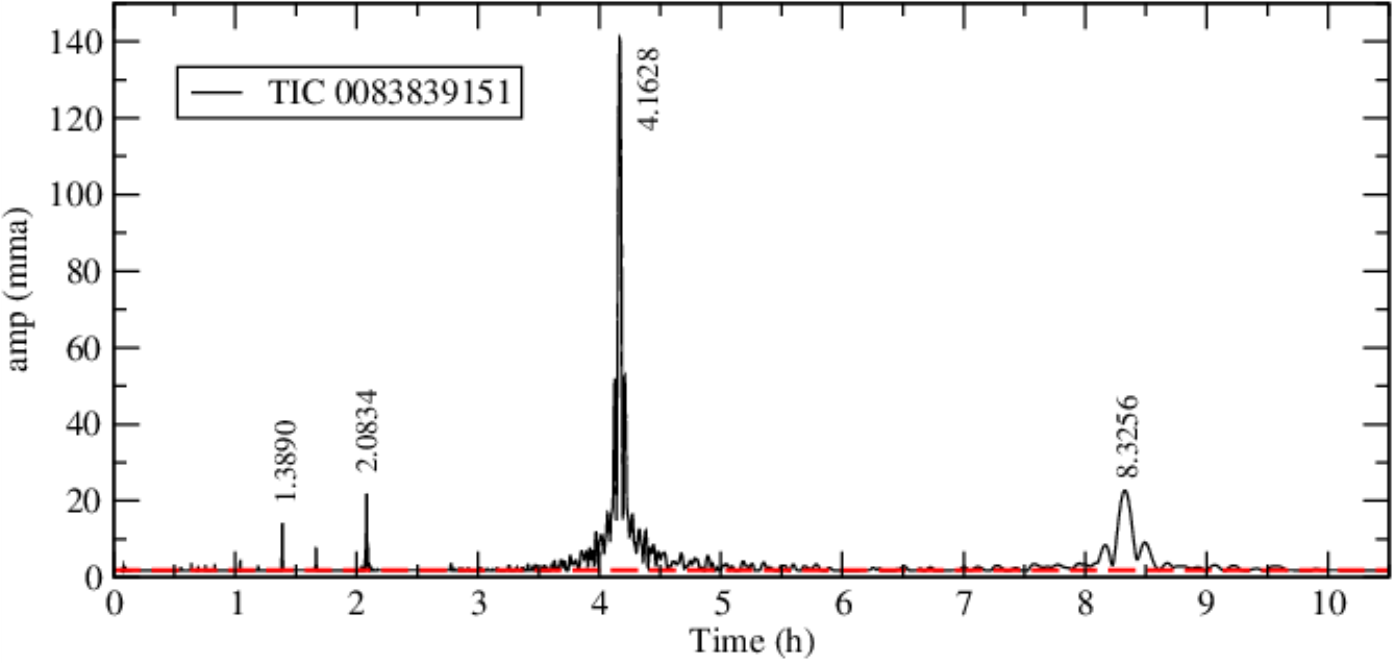}
    	\includegraphics[width=0.85\textwidth]{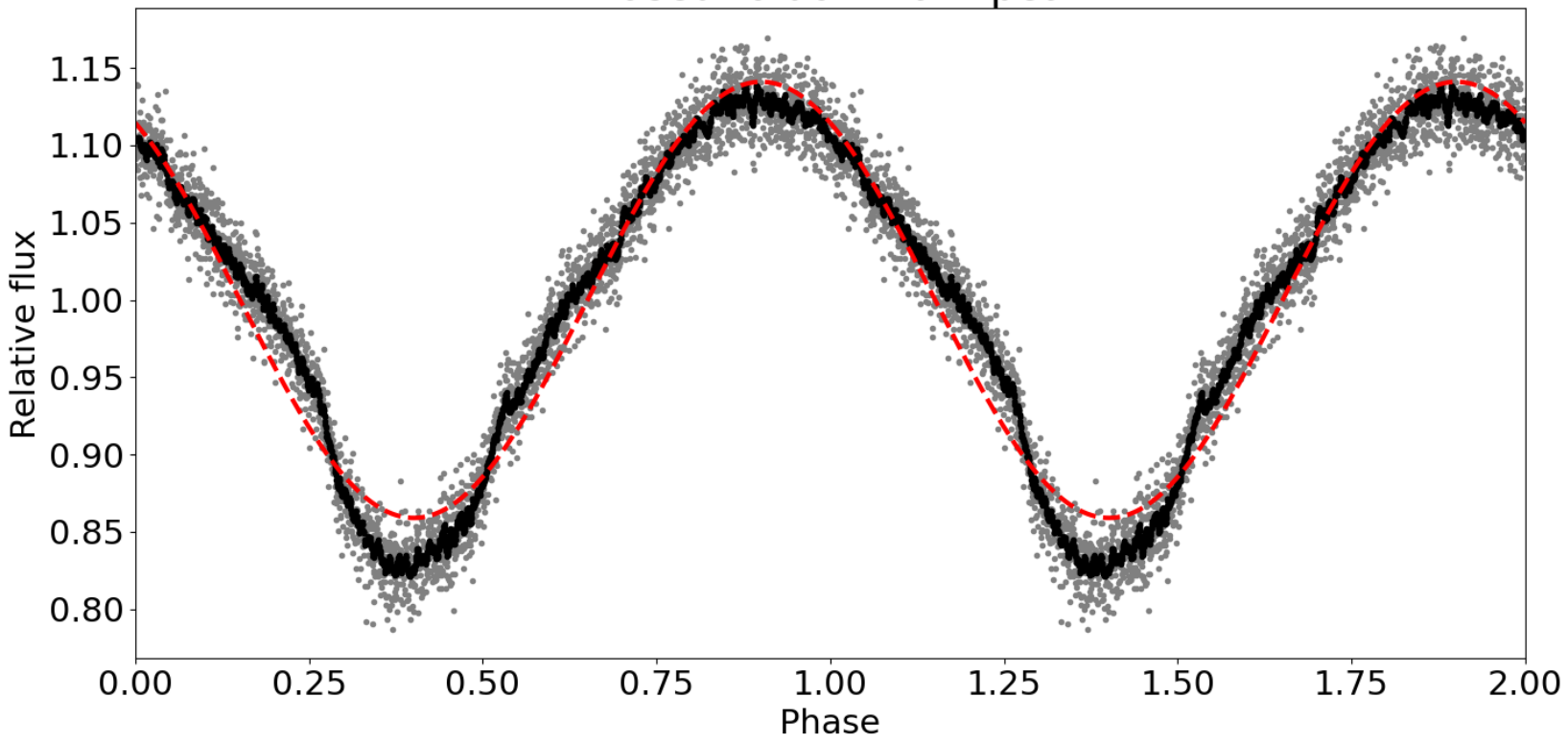}
    	\includegraphics[width=0.85\textwidth]{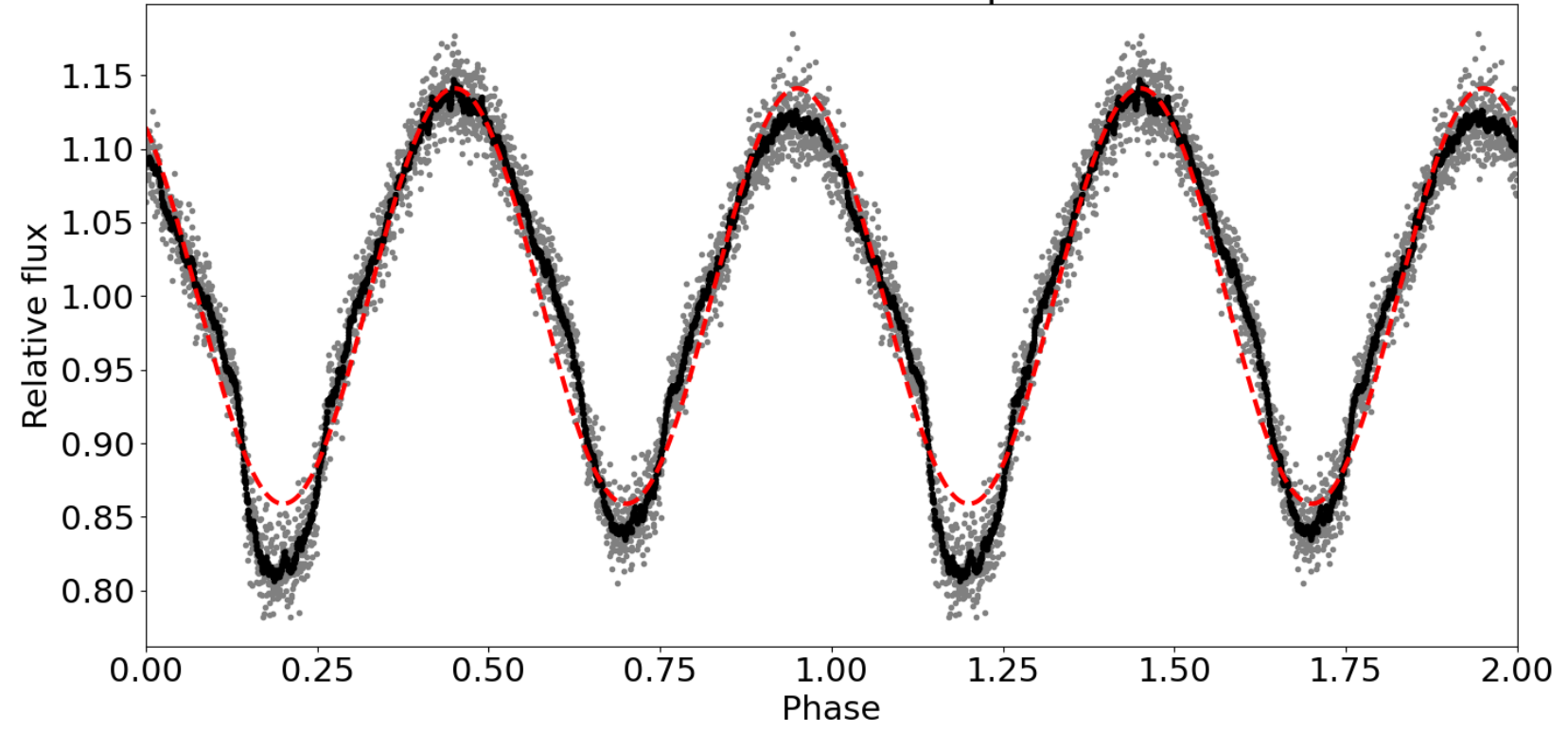}
        \caption{Time series analysis for TIC~008389151. Top panel: Fourier transform of all data. The periods are in hours, and the peak periods are indicated in the figure. The horizontal red line corresponds to the false-alarm probability FAP=1/1000 detection limit. Middle panel (bottom panel): phase-folded light curve over 4.1628~h (8.3256~h). The red dashed line represents a sinusoid with the largest amplitude peak.}
    \label{0083839151}
\end{figure}

For 286 systems exhibiting eclipses {\bf or ellipsoidal variations} in their light curves, the orbital period can be determined directly from these variations. 
As an example of an orbital period that is not associated with the largest peak, we present the data for TIC~0083839151 in Figure~\ref{0083839151}. From the top panel of Figure~\ref{0083839151} in the FT, we see that the period with the highest amplitude is 4.1628~h. The phase-folded light curve over this period is shown in the middle panel of Figure~\ref{0083839151}. However, when we examine the phase-folded light curve over twice this period (bottom panel), we note that the minima are not equally deep, implying an orbital period of 8.3256 h. Note that the period of 8.3256~h is also present in the FT, along with two additional short periods. 

\section{Results}
\label{section3}

In this work, we focus on determining the photometric periods of CVs using TESS photometric data.
The results are presented in Table~\ref{table-results}, which lists the TIC, RA, DEC, FAP (ppt=mma), the detected periodicities, alternative names, literature periods, Gaia DR3 ID, G magnitude, parallax, absolute magnitude, bp-rp colour, CROWDSAP --- the fraction of the object light in the aperture ---, and the TESS-Localize results. The last column includes comments on the system. As mentioned in Section~\ref{section2}, we consider the orbital period to correspond to the peak of the highest amplitude in the FT unless additional information is available in the light curve. We do not attempt to determine the origin of each extra periodicity, as this requires either spectroscopy or the determination of both the orbital and rotational periods to study possible beat periods. In addition, we list the timescales of the long periodicities in the FTs, which might originate from dwarf nova activity or the companion star.
In the comments, we include additional information about the system, such as whether it is eclipsing, whether we detect single or double eclipses, and whether the light curve shows outbursts. For systems classified as magnetic in the literature, we identify whether they are polars or IP CVs. This classification is primarily based on \cite{2023AJ....165..163C}, the \textsc{Simbad} database, the catalogue of polars and candidate polars (VO table, edition 2026) \footnote{https://www.aip.de/de/members/axel-schwope/polarcat/}, and the Koji Mukai  2026 IP catalogue 
\footnote{https://asd.gsfc.nasa.gov/Koji.Mukai/iphome/iphome.html} \citep{MukaiIPCatalog,2025A&A...698A.106S}.

 We determined the optical variability periods for 1362 CVs, selected based on previous classifications as CV. We provide the first determination of the photometric period for 465 objects. Of the 792 CVs with published orbital periods, we found agreement with the literature values for 37\% within 1\%, 55\% within 5\%, 57\% within 10\%, 64\% within 20\%, and 65\% within 30\%. Because of the extended TESS data set, the formal uncertainty in the detected frequencies is on the order of 0.38~$\mu$Hz, arising from the convolution of the $1/T$ and $1/t$ sinc functions, where T is the total data length, and t is the integration time. The observed resolution is determined by the gaps present in the data. In contrast, the literature values are generally based on much smaller datasets and, therefore, have larger uncertainties. CVs show period changes over time, especially after eruptions, and the periods are short, so the numerical differences are significant. We compare our results to those found in the literature in Section~\ref{sec-comparison}. Our results table also lists the literature values for an informed comparison by the reader.

\begin{figure}
    \includegraphics[width=0.86\textwidth]{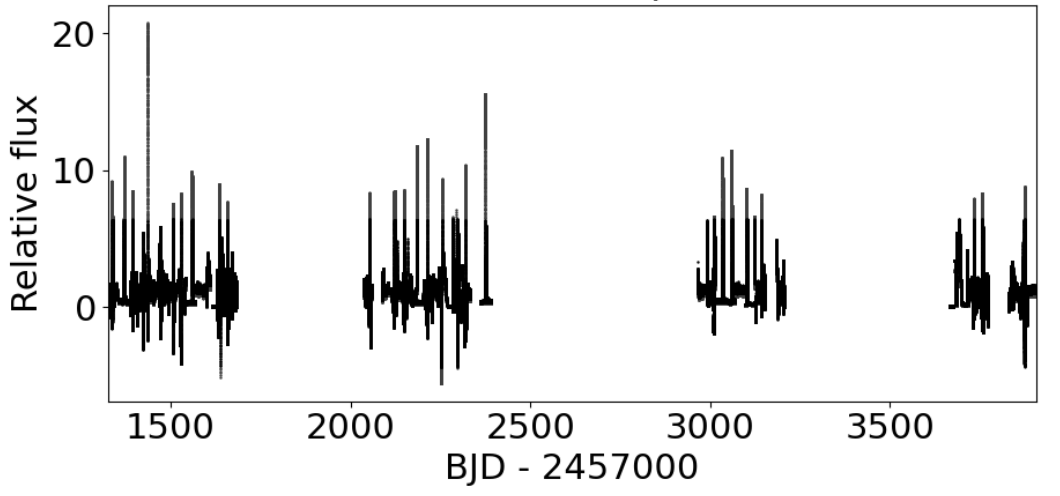}
    \includegraphics[width=0.64\textwidth,angle=270]{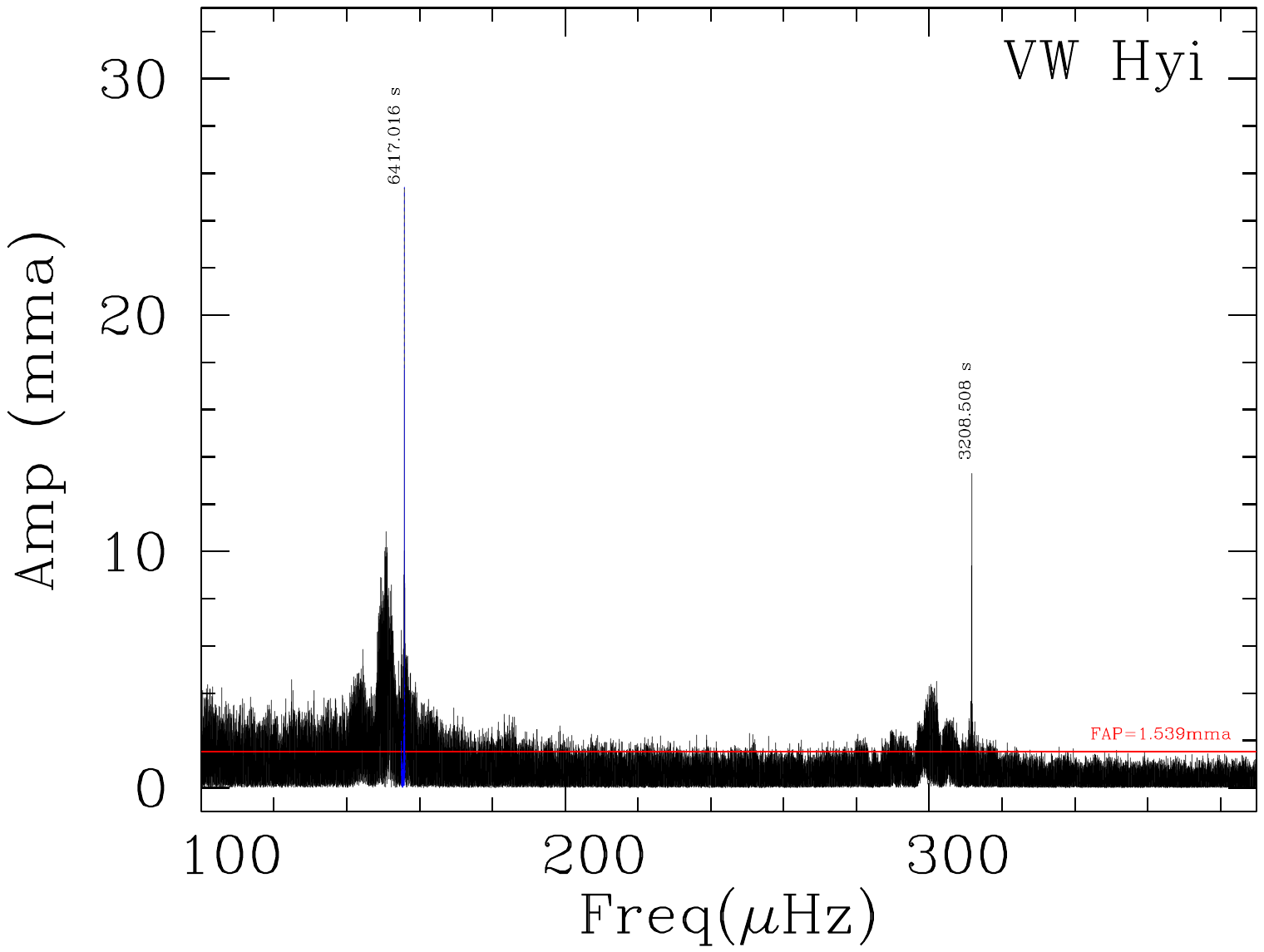}
 \caption{Relative flux as a function of time in BJD for VW~Hyi, TIC~25133286, an SU~UMa-type dwarf novae with G=13.837. It was observed by TESS in sectors 1-13, 27, U29-U37, U39, 61-69, 87-90, 93-98 (U represents 20~s datasets). Note the several peaks indicating bursts that can reach up to 20 times the mean relative flux. The lower panel shows the Fourier transform of the concatenated data in the region of the literature orbital period P=1.7825~h (6417.01~s). The spectral window is shown in blue.} 
    \label{0025133286}
\end{figure}

For 226 CVs, we detected bursting activity in the light curve. As an example, Figure~\ref{0025133286} shows the light curve for the CV TIC~0025133286, VW Hyi, an SU UMa-type dwarf nova, with several peaks in the relative flux corresponding to these events. Many objects show harmonics of the principal peak in the FT, identified as nf$_{\rm i}$ in Table~\ref{table-results}. For most CVs in our sample, we detected additional periods beyond the main period and its harmonics.

\subsection{Discussion of Particular Cases}

\subsubsection{T Leo}
\begin{figure}
\includegraphics[width=0.97\textwidth]{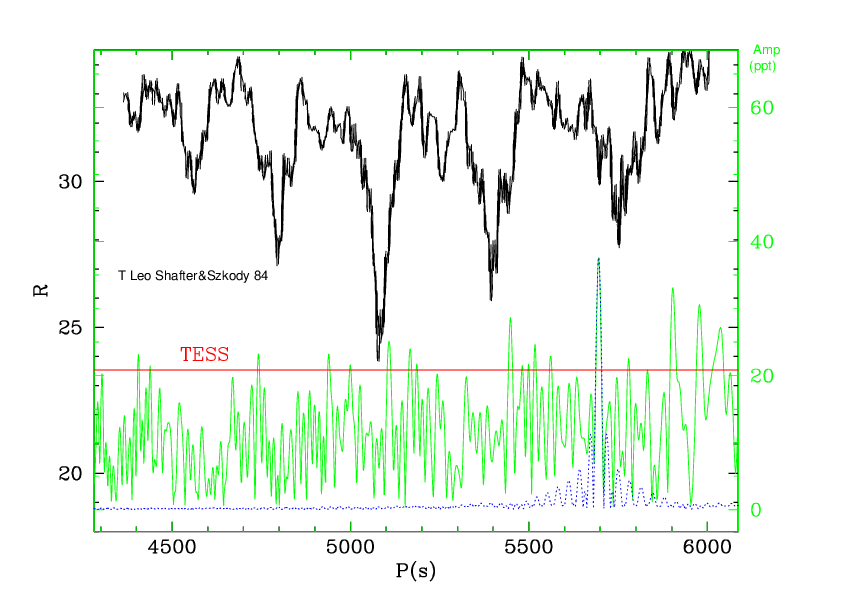}

\caption{Orbital period determination of T Leo from \cite{Shafter84}, their Figure~6 in black, based on 121 radial velocity measurements taken from Jan 4th to April 30th, 1982, with 8~m exposures. The minima of the discriminant R (left vertical axis) indicate the possible periodicities in the data. The most probable orbital period is 1.412~h (5082~s). The two side periods of 1.332~h (4795~s) and 1.500~h (5400~s) correspond to $\pm 1$ cycle-per-day aliases of their most likely period.
    The Fourier transform of the 27~d TESS data, over a single sector but including outbursts, is shown in green (right vertical axis), and the FAP=1/1000 in red. Its main peak, at  $(1.582\pm 0.002)$~h ($5695.8\pm 8.3$~s), corresponds to the second daily alias of the period in \cite{Shafter84}. The spectral window for the TESS dataset is plotted in blue. The uncertainty in period was calculated with a non-linear least-squares fit of a sinusoid to the data.
    \label{tleo}
    }
\end{figure}

One of the systems for which the TESS photometric period differs from the literature value is the old nova QZ~Vir (also known as T~Leo, TIC0396998901). Unlike most stars in our sample, which were observed in multiple TESS sectors, QZ~Vir was observed only in Sector~45 at a 120~s cadence until Sector~102. Consequently, the frequency resolution is lower, and the FT is more susceptible to higher noise than the combined FTs obtained from multiple sectors. \cite{Shafter84} derived an orbital period of 0.058819~d (5\,081.96~s) from 121 spectra obtained between January and April 1982. They also reported two alternative solutions at 0.0555~d (4\,795.2~s) and 0.0625~d (5\,400~s), corresponding to the $\pm 1$ cycle-per-day aliases. Their resulting full width at half maximum (FWHM), including their first daily aliases, is approximately 572~s. Figure~\ref{tleo} compares their period determination (black line) with the FT of the TESS data (green line). The dominant peak in the TESS data occurs at 1.5821~h (5\,695.43~s=0.065919~d), corresponding to the second daily alias of the period reported by \cite{Shafter84}. In contrast, the TESS observations have a FWHM of only 16~s and exhibit aliases with amplitudes of only about 30\% of the main peak on a timescale of 16~d, as illustrated by the spectral window in Figure~\ref{tleo} (blue dashed line). Since QZ~Vir is an accreting system, intrinsic changes in the observed periods over time cannot be ruled out. A Fourier transform using only the 2~d subset of data from $59545.2\leq \mathrm{MJD} \leq  59547.2$, when the CV appears quiescent in the light curve, shows a maximum in the FT at $(0.0594\pm 0.0013)$~d $(5\,136\pm 112~\mathrm{s})$, close to the period in the literature. 

\subsubsection{GK Per}

\begin{figure}
\includegraphics[width=0.99\textwidth]{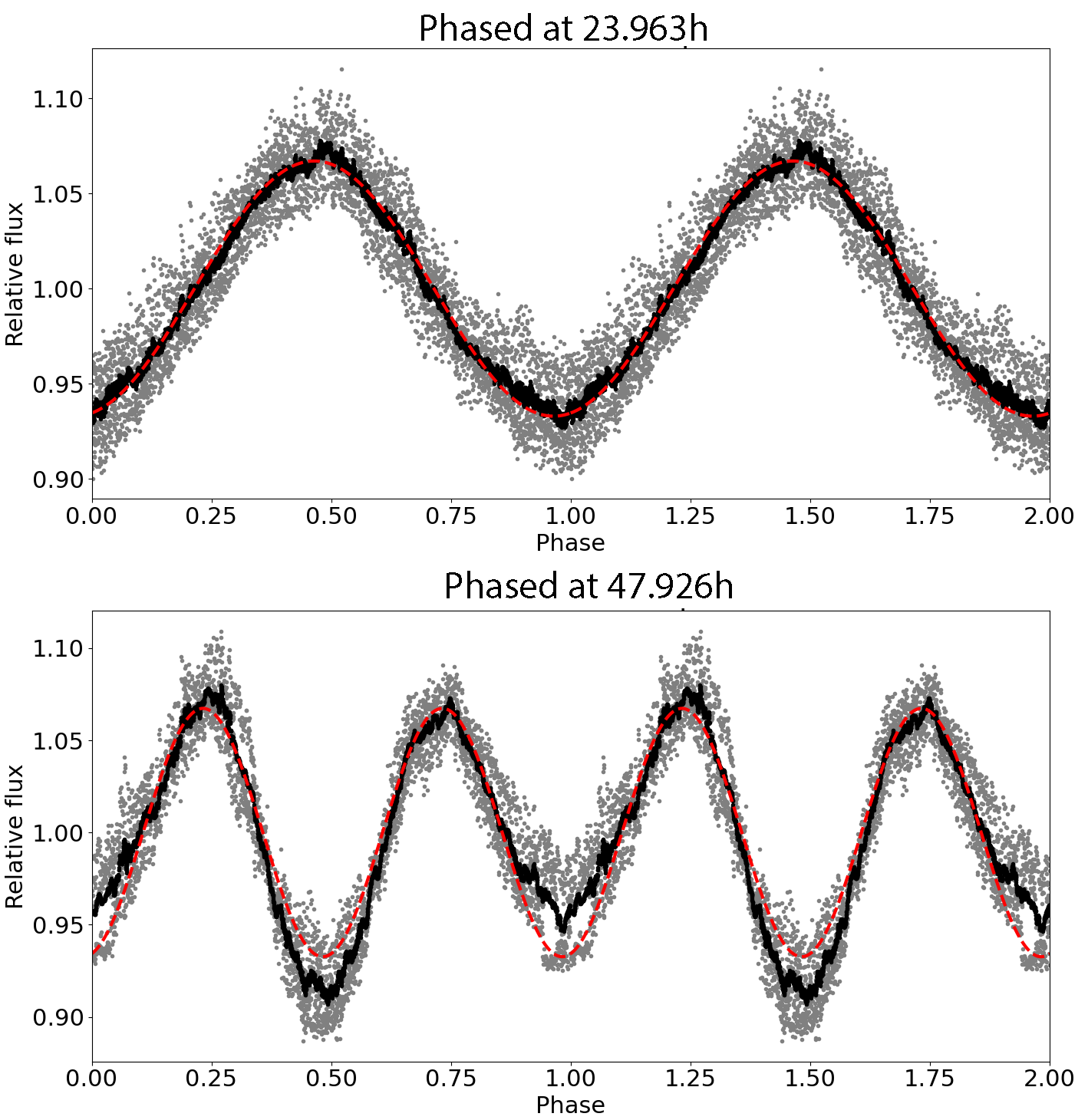}
\caption {Phase-folded light curve for GK~Per from the TESS data. The upper panel shows the light curve folded at the largest-amplitude peak, the first harmonic, while the lower panel shows it folded at twice that frequency. The power spectrum is dominated by a strong signal at twice the orbital frequency and a much smaller one at $\Omega$, mainly due to the ellipsoidal variations of the secondary star.
    \label{gkper}}
\end{figure}

\begin{figure}
\includegraphics[width=0.98\textwidth]{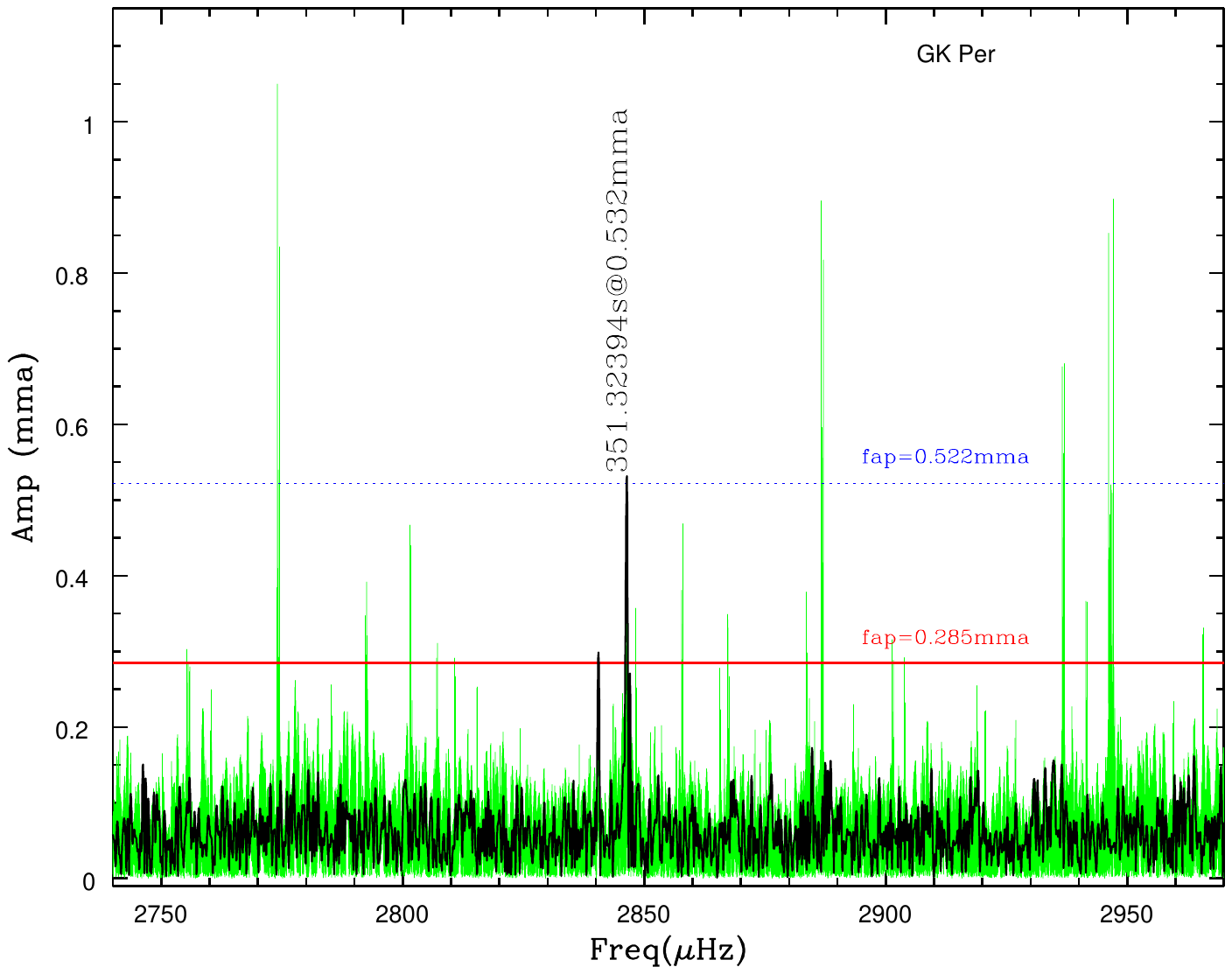}
\caption{Spin period determination of GK~Per from the TESS data. The black curve shows the Fourier transform of the single sector observed at 20~s per exposure. The red line is the corresponding false-alarm-probability FAP(1/1000) line. The green curve shows the Fourier transform of the concatenated light curves from all sectors, the 120~s exposures, and the blue line shows the corresponding FAP(1/1000).
    \label{gkperspin}}
\end{figure}

Another case is GK~Per (Nova~Per~1901, TIC0431762266), an eclipsing intermediate polar observed by TESS in Sectors 18, 58, and 86 at a 120-s cadence; Sector~58 is also available with 20-s exposures. Its orbital period of almost 2 days ($1.996872\pm 0.00009$~days) makes it difficult to observe all phases from the ground. The orbital period measured from the TESS data is consistent with the literature value, $P_\mathrm{orb}=\Omega=172,454.0$~s (47.9~h) \citep{AH21}, whereas the spin period ($\omega$) is more accurately determined from the higher-cadence observations. The FT of the three concatenated 120-s light curves is dominated by the orbital modulation at 171,988.22~s (47.77~h) with an amplitude of 24.11~ppt, together with its first harmonic at 86,265.13~s (23.962~h), which reaches an amplitude of 67.24~ppt (Figure~\ref{gkper}). A weaker signal is detected at 360.49~s with an amplitude of 1.05~ppt, above the false-alarm probability threshold of 0.52~ppt; however, additional peaks are likely due to instrumental effects, amplitude fluctuations, or stochastic flickering. The spin period of 351~s in X-rays, which established GK Per as an IP, was reported by \citet{Watson85}. The literature orbital and spin periods are 172,454.0~s (47.9~h) \citep{AH21} and $351.332\pm 0.002$~s \citep{Mauch04}, respectively. The 120~s TESS data result in a photometric period of $(47.904\pm 0.015)$~h ($172\,454\pm 52)$~s. The 20-s observations from Sector~58 provide both a higher Nyquist frequency and improved frequency resolution for the short-period signal, but not for the long-period signal because the data set is much shorter. In these data, we
%we measure an orbital period of 179\,518\pm 811~s (multif) (FFT 172,291)~s (49.866~h) and 
recover the spin period at $351.33\pm 0.20$~s, in excellent agreement with the published value (Figure~\ref{gkperspin}). Because GK~Per undergoes eruptions and ongoing accretion, small secular changes are observed. 
\cite{Bruch25} detected in sector 18 $\omega=351.3209\pm 0.0097$~s, for sector 58 $\omega=351.3241\pm 0.0041$~s, plus $\omega-\Omega$ and 2$\omega$.


%\subsubsection{RS~Oph}

%Another challenging case is the symbiotic star RS~Oph --- not a CV --- whose orbital period is 453.6~d. TESS observed the system in only a single 27-day sector (through Sector~102), making it impossible to recover such a long period from the available data. The Fourier transform of the TESS light curve reveals several significant periodicities above the false-alarm threshold of FAP(1/1000)=2.21~ppt, with the strongest peaks at 1.66~d (44.28~ppt) and 0.95~d (41.36~ppt), along with several lower-amplitude signals. None of these corresponds to the known orbital period. Five Gaia sources are located within 15 arcsec of RS~Oph. This is a rare case in which TESS-Localize distributes the likelihood across all five stars and cannot uniquely identify the source of the detected variability. However, the TESS CROWDSAP value of 0.959 indicates that 95.9\% of the flux within the aperture originates from RS~Oph. Although contamination from nearby stars cannot be completely ruled out, the detected light is strongly dominated by RS~Oph. For completeness, we include the detected periods in our catalog and note the ambiguity in the source localization.

\subsubsection{AT Cnc}

\begin{figure}
\includegraphics[width=0.70\textwidth,angle=270]{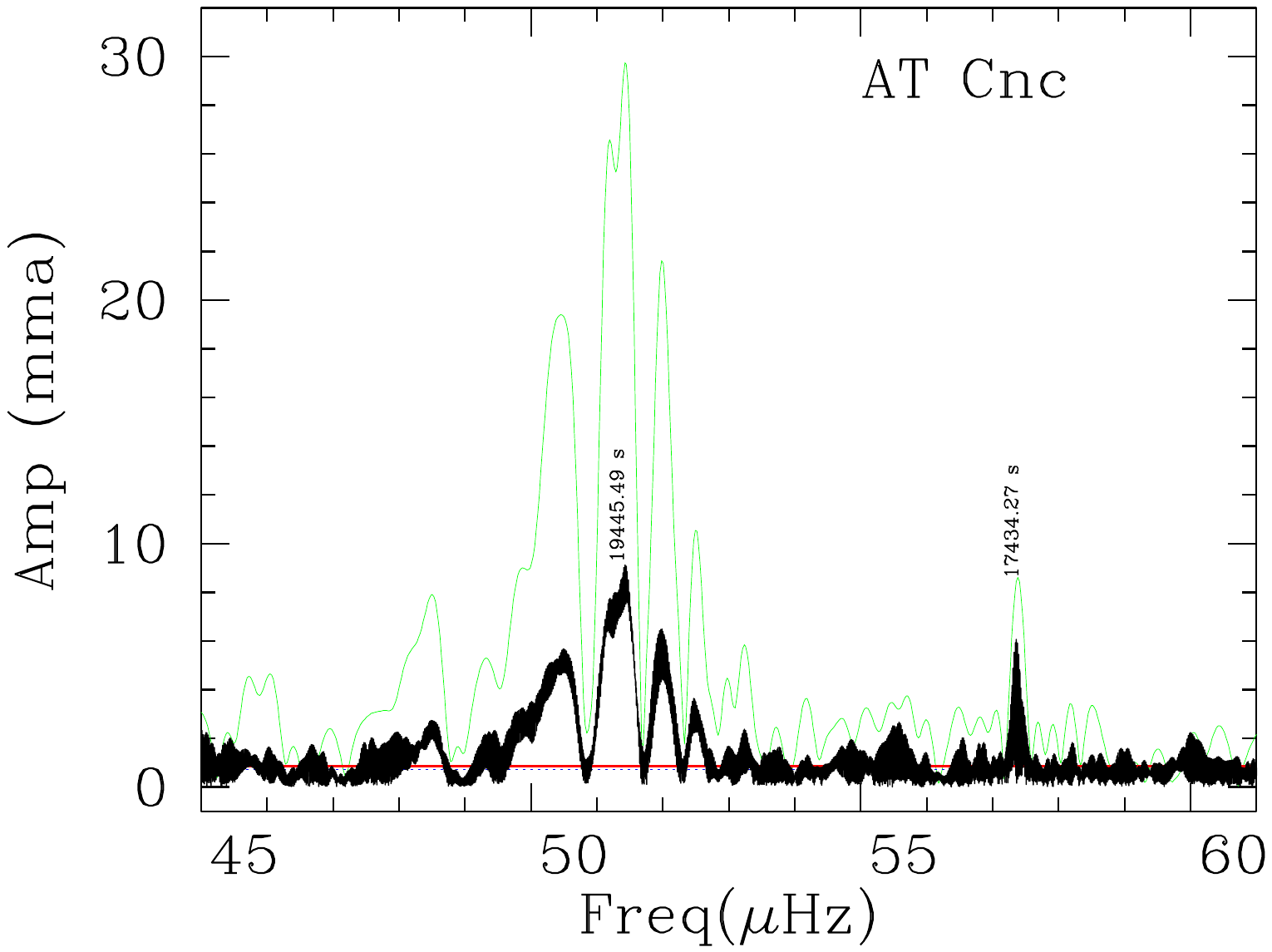}
\includegraphics[width=0.55\textwidth]{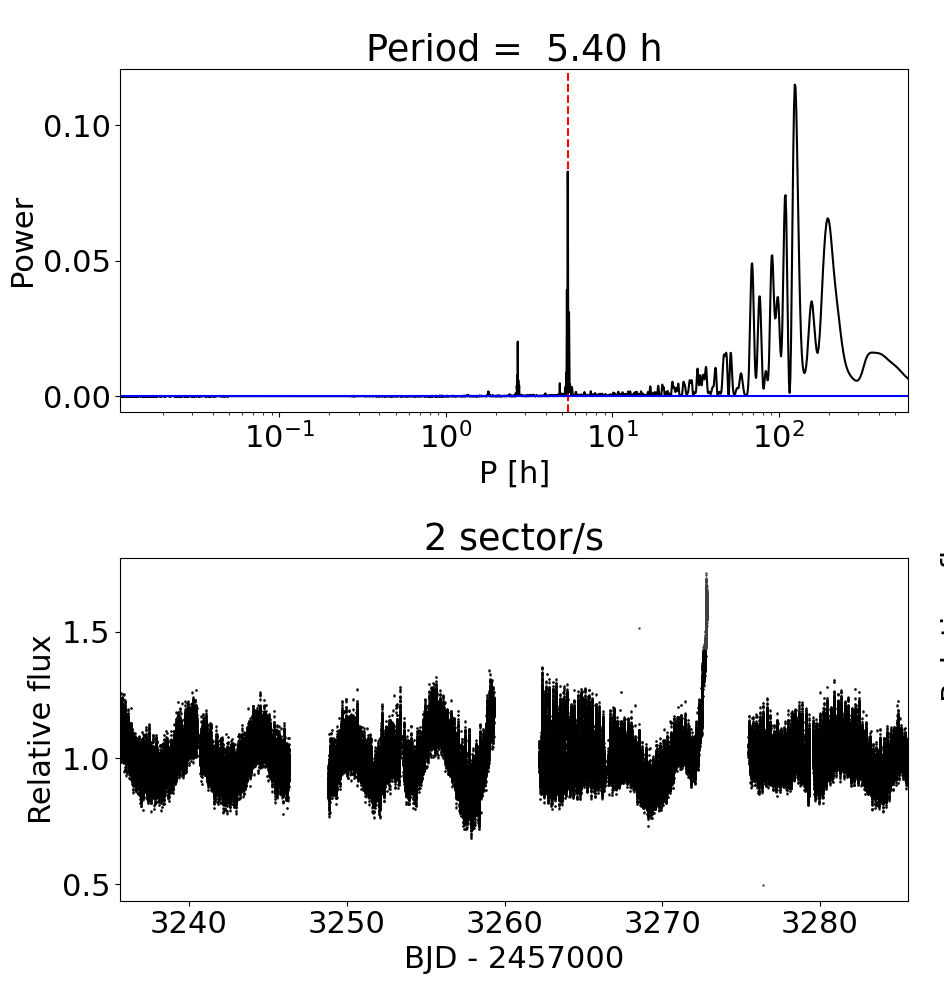}

\caption{Fourier transform, periodogram, and light curve for the two 20~s-per-exposure sectors 71 and 72 of AT~Cnc. The largest-amplitude peak in the periodogram is from the superhumps, with a period of 5.4014~h (19445~s), which is visible in the light curve. The FT shows both the 120~s data (in black) and the 20~s data (in green). The orbital period is 4.8428~h (17434~s). The FAP(1/1000)=0.86~mma is shown as a horizontal blue line.
    \label{atcnc}}
\end{figure}
For the IP AT~Cnc, TIC~3645653,
a very active system that hardly ever remains in quiescence but exhibits a rapid succession of outbursts, 
\cite{Nogami99} obtained $P_\mathrm{orb}=17375.0$~s and $P_\mathrm{spin}=1603$~s from time-resolved spectroscopic observations obtained during a standstill. They measured an orbital period of (4.83$\pm$0.01~h) from low-amplitude photometric variations, with possible periods of 0.249~d (21514~s) or 0.132~d (aliases thereof) in their light curves. This is different from the period of 0.239~d=20650~s claimed to be present in AT~Cnc by \cite{Gotz86}, and none is close to the orbital period.
\cite{Kozhevnikov04} optical photometry could not confirm the variations found by \cite{Gotz86} and \cite{Nogami99}, but by separating his data into two sets adjacent in time, he instead finds indications of quasi-periods of 4.65 and 4.74~h, slightly less than the spectroscopic orbital period, with amplitudes of 5–9~mmag. He interprets these variations as negative superhumps, i.e., caused by brightness variations due to the nodal precession of an accretion disk inclined with respect to the binary's orbital plane.
\cite{Kozhevnikov04} also detected a broad hump at frequencies of 0.4–0.7~mHz (corresponding to periods between $\simeq$25–40 min) in the power spectra of his light curves.
\cite{Bruch19} obtained 37 light curves during two episodes of standstill in AT~Cnc in 
February – April 2016 and March – April 2018. 
The average of the corresponding periods in the two
seasons is 4.839~h (17\,420~s). Within the error bars quoted by \cite{Nogami99}, it is identical to
the spectroscopic orbital period, lending credibility to the reality of this photometric period, which they identify with the orbital period. The uncertainty was 0.008~h. Because of the 2~yr gap, \cite{Bruch19} quoted
$P_\mathrm{orb} = 0.201634$~d (4.83922~h)
or the alias $P_\mathrm{orb} = 0.201580$~d (4.83792~h = 17417~s), and
$P_\mathrm{rot} = 26.73161\pm 0.00007$~min ($1603.8966 \pm 0.0042$~s).

For the TESS data, the 120~s cadence observed in sectors 21, 44-47, and 71-72 has FAP(1/1000)=0.86~mma, showing periodicities at $(5.40148\pm 0.00014)$h=$(19\,445.31\pm 0.51)$~s@14~mma,
2f1=$(9741.46\pm 0.37)$~s@5~mma + 3 harmonics, which \cite{Bruch25} interpreted as superhumps, 
$(4.8428\pm 0.0002)$~h=$(17434.12\pm 0.80)$~s@7~mma, the literature orbital period, and other smaller-amplitude periodicities, like
$(1415.70\pm 0.03)$~s@1~mma.
%103569.21~s@9.804mma, 146349.19~s@15.88mma, 197347.155~s@21.61mma, 
Two outbursts occurred around BJD=1850. For the 20~s light curves 
U71-U72, fap=0.746~mma, the superhump period is 
$(19434.9\pm 4.9)$~s@32~mma,
with a harmonic at 
$(9743.5\pm 2.4)$~s@16~mma,
and the orbital period is 
$(17431.0\pm 25.1)$~s@5~mma.

Unlike our approach of concatenating all sectors after normalizing each one individually, \cite{Bruch25} studied each sector separately.

\subsubsection{V379~Vir}
The TESS data for V379~Vir, SDSS~J121209.30+013627.7, TIC0953321496, with
G=17.984, observed in 
sectors 46 and 91 at 120~s cadence, with CROWDSAP:0.852, show only two periods below 28~h: $(470.7996\pm 0.0006)$~s at 11~mma and $(459.2495\pm 0.0006)$~s at 10~mma, close to the FAP=(1/1000) of 10~mma. 
It was classified as a DAH+M by \cite{Silvestri07}. \cite {Amorim23} determines a magnetic field of B=11~MG for the DA white dwarf. \cite{MunozGiraldo24} shows an L8 donor with a mass of $0.05\,M_\odot$ and a
magnetic white dwarf with $P_\mathrm{orb}=5298$~s, $T_\mathrm{eff}^\mathrm{wd}=10\,000$~K and mass $0.640\,M_\odot$, while \cite{Breedt12} classifies it as a Polar with weak H$\alpha$ emission but no evidence of a companion star. \cite{Suslikov25} estimates  $M_\mathrm{wd}=0.61\pm 0.05\,M_\odot$,  $T_\mathrm{eff}^\mathrm{wd}=10\,930\pm 350$~K, and for the brown dwarf radius $R=0.095\pm 0.018\,R_\odot$ and $T_\mathrm{eff}=1600\pm 180$~K. They estimate a $dM/dt=3\times 10^{-13}M_\odot$/yr, too low for a CV, where the mass transfer rate $dM/dt \geq 10^{-11}\,M_\odot$/yr is expected, while for low accretion rate polars (LARP) $dM/dt \geq 10^{-14}\,M_\odot$/yr. 
In LARPS, the infalling material does not form
a shock but directly heats the atmosphere
of the white dwarf to temperatures of $kT\sim 1$~keV, and 
cooling occurs primarily via cyclotron radiation \citep{Kuijpers82,Woelk92}.
\cite{Schmidt05} classifies it as a CV with a magnetic white dwarf with B=7~MG plus an L dwarf, and \cite{Debes06,Farihi08} detected variable cyclotron emission with a period of $88\pm 1$~min ($0.061\pm 0.001$~days) and B=7~MG, while \cite{Stelzer17} with X-ray XMM observations shows that the variability is modulated at $P_\mathrm{orb} = (88.3\pm 0.6)$~min, with a bin size of 450~s. The orbital period was confirmed with optical photometry at $5305.660\pm 0.006$~s, obtained with ULTRACAM and 10-s exposures by \citet{Burleigh06}. The TESS light curves do not detect the known orbital period above its FAP(1/1000). A multisinusoid fit with the known orbital period results in $(5305.58\pm 0.09)$~s at 9~mma. The short photometric periods detected, even if confirmed, are not orbital.

\subsubsection{1ES 1210-646}
TIC0382408105, also called 1ES~1210-646 and 4U~1210-64, 
is listed in \cite{Revnivtsev07,2023AJ....165..163C} as an IP based on its X-ray spectrum. The orbital period
 $P=6.698\pm 0.020$~d was calculated by \cite{Corbet08,Coley14} from X-ray variability.
It was classified by
\cite{Masetti09,Masetti10} as a high mass X-ray binary, while \cite{2023AJ....165..163C} classifies it as a non-magnetic CV.
It was reclassified as a X-ray binary by \cite{Bruch25}, and observed by TESS in sectors 64-65 and U99-U100 (20~s), showing $P_\mathrm{orb}=(6.8772\pm 0.0008)$~d and a periodicity of $(3.52674\pm 0.00011$)~h [$(12\,696.26\pm 0.4)$~s].

\begin{figure}
\includegraphics[width=0.99\textwidth]{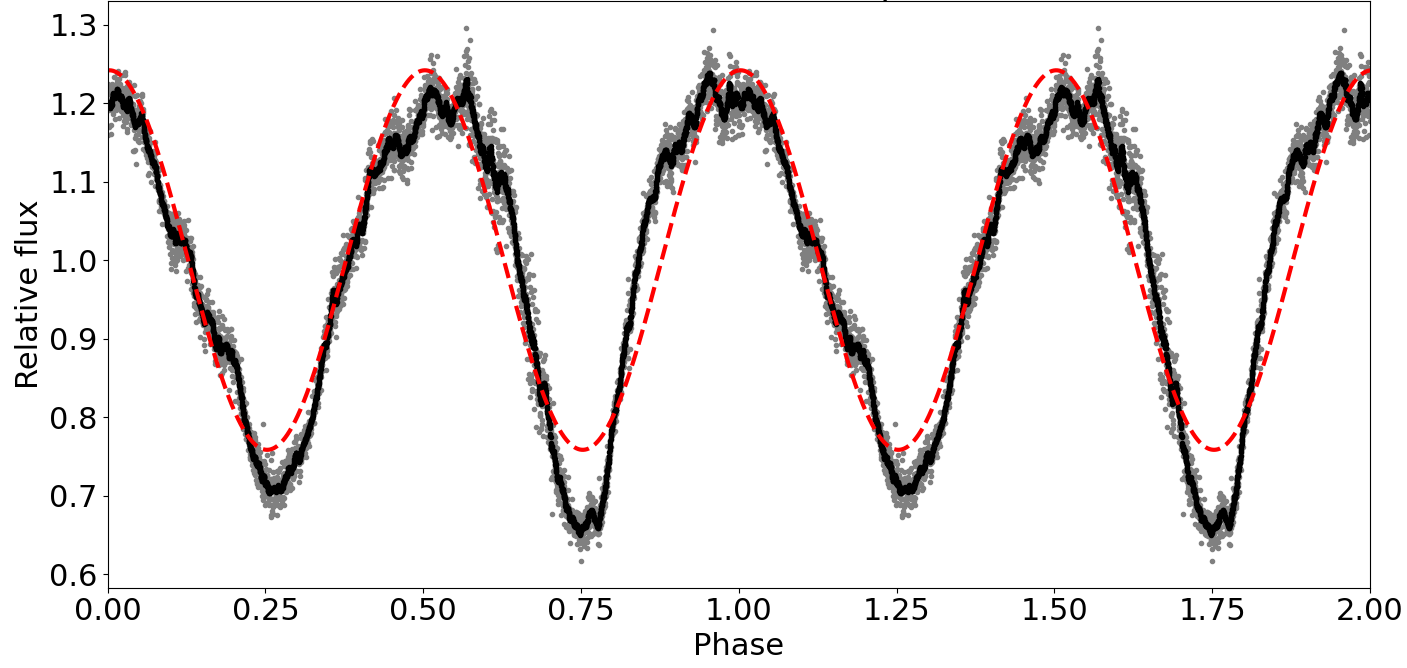}
\caption{Light curve for 1ES~1210-646 for sectors 64--65 and 99-- 100, folded at a period of 6.8772~d.
    \label{1es}}
\end{figure}

\subsubsection{Tau 4}

\begin{figure}
\includegraphics[width=0.70\textwidth,angle=270]{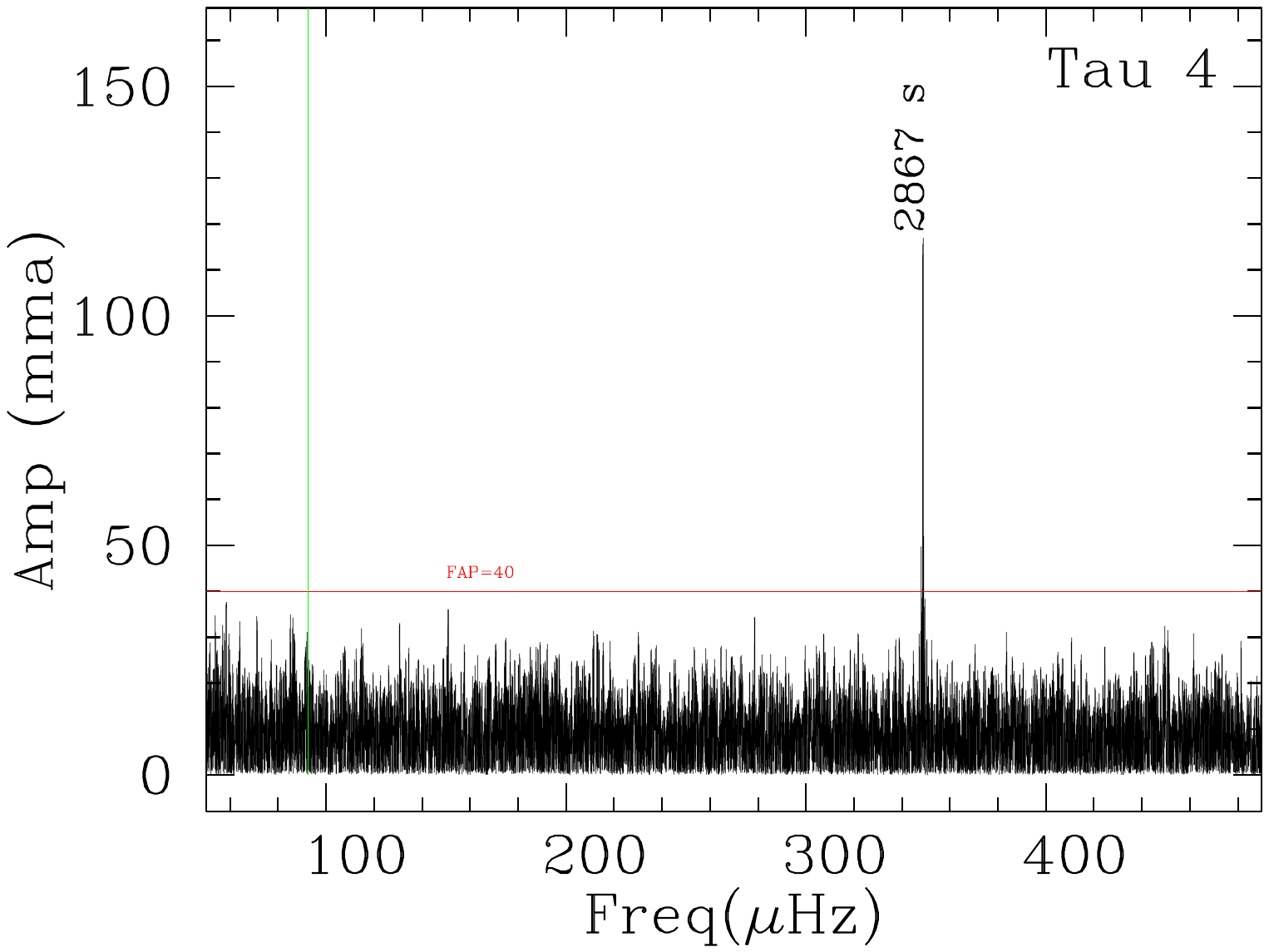}
\caption{Fourier transform of the TESS light curve for Tau~4. The green vertical line indicates the position of the spectroscopically estimated orbital period from \cite{Howell08}. 
    \label{tau4}}
\end{figure}

[DWS97]Tau~4, also called RX~J0502.8+1624 and TIC303680865, is a polar published by \cite{Howell08}, with B=7--11 MG and an orbital period $P_\mathrm{orb}=10\,800$~s derived from spectroscopy. Their photometric light curve shows no periodic signal. It is a faint source with G = 18.847.
The TESS data from sectors 71 and 98, with CROWDSAP:0.017, indicate strong contamination from nearby stars
and show a photometric period of $(2866.71\pm 0.26)$~s at 137~mma, and a FAP(1/1000) = 40.0~mma. 
TESS-Localize indicates that the signal is coming from the CV, with p-value = 0.196, likelihood = 1, and height = $0.746\pm 0.054$. There is no peak around 10\,800~s above the FAP (see Figure~\ref{tau4}). A multisinusoid fit with the orbital period results in $(10\,782\pm 18)$~s at $(27\pm 17)$~mma.

\subsubsection{Values in the literature\label{swrong}}
The periods published for the stars listed in Table~\ref{wrong}, based on TESS light curves, do not come from the cataclysmic variable system but from nearby stars.

\begin{table*}
\scriptsize
%\begin{deluxetable*}{|l|l|l|}
\caption{Contamination from nearby stars\label{wrong}}
%\startdata
\begin{tabular}{|l|l|l|}
\hline\hline
TIC & Name &Source, Period, Information \\
\hline
0004970499 & 2MASSJ09275195-3910524 & \cite{Dag26} P=19467.8~s, double eclipse from 35" 2MASSJ09275494-3910527 5429765619165769600 G=12.415 EB gaivar \\
0040223843 &V823Cyg &\citet{Dag26} P=16308~s, from 2059533266224998784 EB \\
0159580213 & KIC8751494 & \citet{Dag26} P=31032~s, 31119.57~s double eclipse from 30" EB ATO J291.0335+44.9915 Gaia DR3 2126637221076538624\\
0267353881 & CRTSJ173516.9+154708 & \citet{Dag26} P=49740.9~s, correct 30477.97s@21.91mma, 2f1=15235.49~s@84.39mma \\
0620724990 & SDSSJ015051.52+332621.8 & \citet{Dag26} P=18000s, 17982.44~s from Gaia DR3 305760199271364096 G=12.128312 \\
0741679184 & EQLyn  &\citet{Dag26} 18036~s from 59" 927255852632972288 ATOJ116.3626+45.6335 RSCVn G=13.594 gaiavar\\
1140118632 & CXOPSJ154305.5-522709 & \citet{Bruch25,Hernandez25} P=15652.224~s, from Gaia DR3 5886078406583748352 G=19.550 \\
1211418907  & NRTrA& \citet{Bruch24a,Dag26} P=47\,234~s,  double eclipse from 24" Gaia DR3 5831361008602356224 eclipsing binary G=13.431 \\
2001466142 &  ASASSN-14cl & \citet{Dag26} P=49\,680~a, double eclipse from Gaia DR3 1796892927988167424 G=14.644 \\
\hline
\end{tabular}
\end{table*}
%\enddata
%\end{deluxetable*}

\subsection{Period distribution for CVs}

In a cataclysmic variable, the low-mass main-sequence secondary star transfers mass via Roche-lobe overflow. For mass transfer to continue, the binary orbit must continuously shrink so the secondary remains in contact with its Roche lobe.

Magnetic braking is a fundamental driver of cataclysmic variable (CV) evolution, as it facilitates the angular momentum loss (AML) necessary for these systems to progress through different orbital phases. This process occurs when ionized particles from the stellar wind are forced to follow magnetic field lines, creating a lever arm that efficiently extracts angular momentum from the rotating secondary star \citep[e.g.][]{Webbink02}. 

At orbital periods longer than 3 hours, the primary mechanism driving the orbital contraction is magnetic braking. For shorter periods, the star becomes fully convective, and magnetic braking drops drastically.
It halts mass transfer until gravitational radiation drives the system’s components back into contact, allowing accretion to resume and thereby determining the location and width of the period gap \citep{Howell01,Knigge11,Zhou26}.

\begin{figure}
\includegraphics[width=0.97\textwidth]{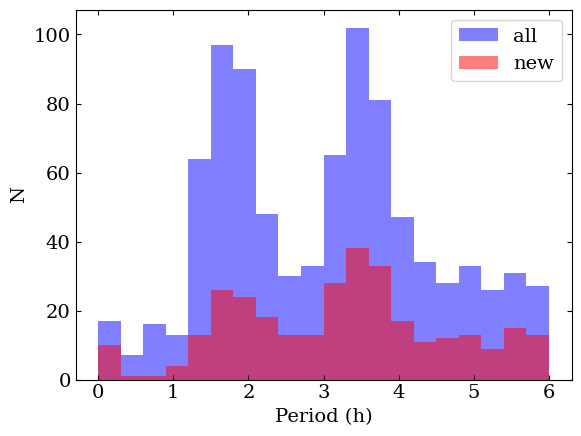}
    \caption{Distribution of photometric periods of the 1000 CVs (73\%) from our sample with periods up to 6\,h. The period values obtained in this work using TESS data are shown in red, while those from the literature are shown in blue.
    \label{TESS-12hs}
    }
\end{figure}

In Figure~\ref{TESS-12hs}, we show the main photometric period distribution for all CVs in our sample with periods up to 6\,h (blue). This amounts to 1004 objects, which is 74\% of the total sample of 1362 CVs. With the caution that the sample does not represent an unbiased population (see Section~\ref{section2}), the distribution shows the expected period gap between $\sim$2 and $\sim$3\,h, but this gap is not entirely devoid of objects. In our sample, 147 CVs (11\%) have periods between 2.15~h and 3.18~h, which fall within the period gap range defined by \citet{1983ApJ...275..713R}.
Note that the period distribution of the CVs with new periods (red) closely matches that of the total sample.
Systems with orbital periods below $\sim$70~minutes should have compact companions rather than main-sequence companions due to the smaller orbital period \citep{2025A&A...700A.107G}. In our sample, we have 53 CVs with periods shorter than 70 minutes. A small fraction of those short-period systems may be IPs whose orbital modulation remains undetected. Finally, 25\% (299 CVs) of the sample show periods below the gap, while 64\% (767 CVs) show periods above 3.15~h. The median period for the complete sample is 3.6738\,h.

%\subsection{novae}

\begin{figure}
    \includegraphics[width=0.97\textwidth]{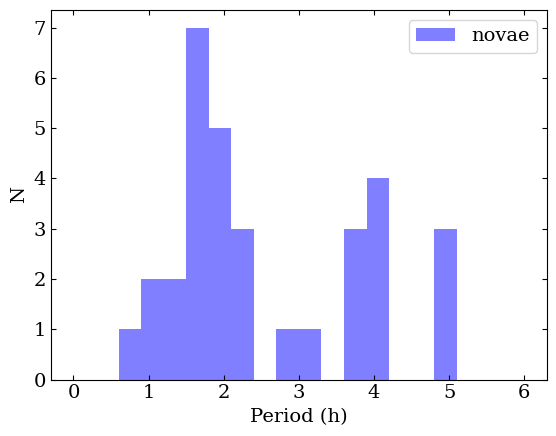}

    \caption{
    Photometric period distribution for novae, up to 6\,h. }
    \label{novas-12hs}
\end{figure}

The detection and classification of novae are subject to strong observational bias: they are often classified as CVs upon detection of an eruption; consequently, the period distribution of these objects is biased \citep{1997A&A...322..807D}. They also suggested that the period distribution of classical novae does not exhibit a period gap; instead, it is much more uniform.  Our sample has 44 CVs classified as novae from the literature. Figure \ref{novas-12hs} shows the period distribution of the 32 objects with photometric periods shorter than 6\,h. 
Note that most objects have short periods, mainly between 2 and 5\,h, which contribute to the peaks both below and above the known period gap.  The median of the distribution of novae is 12\,104~s, or 3.362~h. Interestingly, the short-period, low-angular-momentum-loss systems (<3 h) outnumber the longer-period ones.
Given the small number of novae compared to the total sample, we do not expect our period distribution to be strongly influenced by novae.

\subsection{Eclipsing systems}

\begin{figure}
    \includegraphics[width=0.97\textwidth]{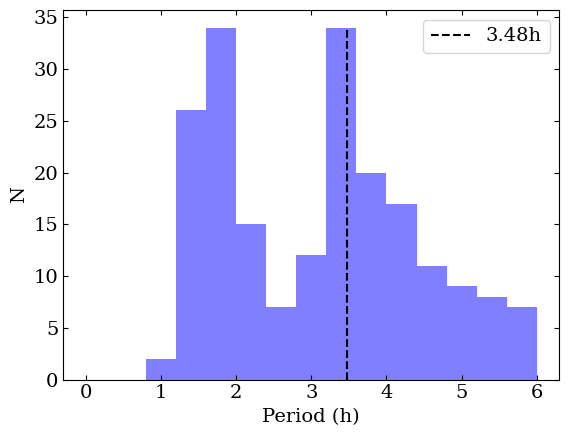}

    \caption{Distribution of periods for the 205 CVs classified as eclipsing, or showing ellipsoidal variations, with orbital periods up to 6~h, with a median of 3.48~h. }
    \label{TESS-eclipse-10hs}
\end{figure}

For light curves showing eclipses or ellipsoidal variations, the orbital period is straightforward, as mentioned in Section~\ref{section2}. Our sample includes 286 CVs with visible eclipses or ellipsoidal modulation in their light curves; therefore, {\bf we are confident that these represent orbital periods}. Most orbital periods for eclipsing systems range from $\sim$2~h to 15~h, with only five objects having periods longer than 15~h. Figure~\ref{TESS-eclipse-10hs} shows the period distribution for the 205 eclipsing CVs with periods of up to 6\,h. In this sample, we observe a distinct period gap between $\sim$2.5 and 3\,h, consistent with the results for the complete sample shown in Figure~\ref {TESS-12hs}. The median value for the sample of 286 eclipsing or ellipsoidal variation CVs is 3.481~h (12530~s). We are aware that ellipsoidal variations are more readily observed in long-period systems, but these periods should be orbital.

\subsection{Magnetic CV}
\label{sec3.5}

In this section, we focus on magnetic CVs, i.e., polars ($B\gtrsim 10$~MG) and intermediate polars (IPs), in which the magnetic field ($B\gtrsim 0.1$~MG) influences mass accretion onto the white dwarf. Polar CVs are systems in which the magnetic field of the white dwarf component is strong enough to prevent the formation of an accretion disk \citep[e.g.]{Zhou26}. In this case, mass accretion occurs through the white dwarf's magnetic poles. It is assumed that the magnetic field keeps both stars in synchronous rotation \citep{2004ApJ...614..349N}; thus, the white dwarf's rotation period equals the binary's orbital period within a few percent \citep{2025A&A...698A.106S}. 

\begin{figure}
    \includegraphics[width=0.50\textwidth]{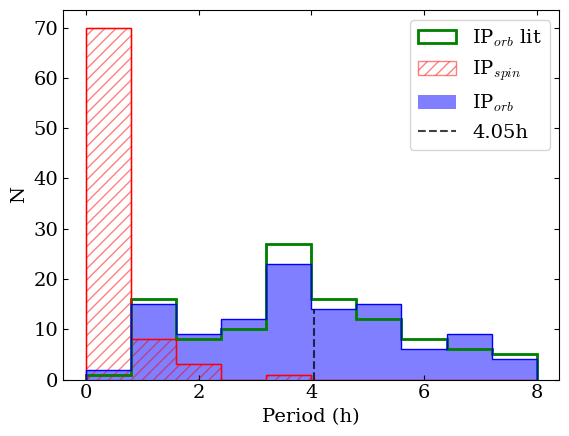}
    \includegraphics[width=0.50\textwidth]{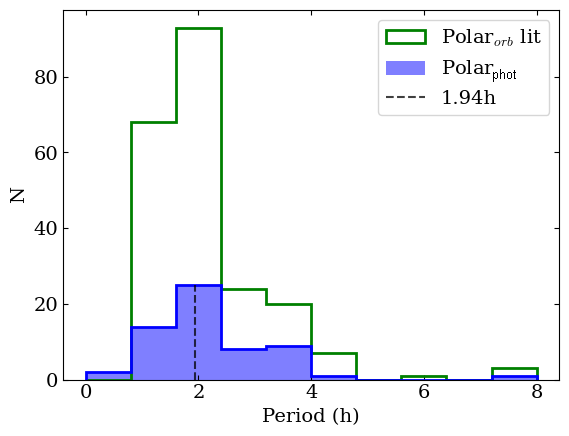}
    \includegraphics[width=0.50\textwidth]{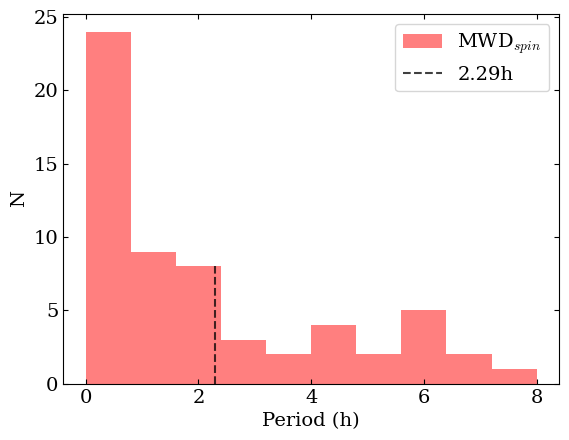}
        \caption{Period distribution for the sample of IPs (upper), polars (middle), and single magnetic white dwarfs (bottom), with periods up to 8\,h. The vertical dashed line indicates the median value of the period distribution for the sample presented in this work. For the IPs and polars, we include the period distribution from the literature. 
        }
    \label{magnetic}
\end{figure}

From our sample, we identified 63 polar CVs, classified as such in the literature, all with orbital periods below 8~h.  The photometric period distribution for the 63 polar CVs is depicted in blue in the middle panel of Figure~\ref{magnetic}. As shown in the figure, the period distribution of polar CVs differs from that of the CV population (see Figure~\ref{TESS-12hs}). A period gap is not expected because, for polars, the white dwarf's magnetic field reduces the donor star's wind zones, thereby reducing the efficiency of magnetic braking \citep{2020MNRAS.491.5717B,2024A&A...682L...7S}. However, its presence cannot be ruled out in our sample distribution. The median period of the 63 polar CVs analyzed in this work is 6904~s (1.92~h).

For intermediate polars, the magnetic field of the white dwarf is weaker (B$\lesssim$10~MG) than that of polars (B$\gtrsim$10~MG) \citep[see][for example]{2023MNRAS.524.4867I}. For IPs, the magnetic field is not strong enough to force the white dwarf's rotation period to synchronize with the system's orbital period. For most IPs, the spin period of the white dwarf is significantly shorter than the orbital period, by a factor of around 10 to 100$\times$ \citep{2004ApJ...614..349N}. 
We identified 136 CVs as IPs in our total sample. The orbital period distribution for the 108 (79\%) IPs with orbital periods of up to 8~h is shown in the top panel of Figure~\ref{magnetic} (blue). Up to 6~h, there are 92 IPs (68\%).

The orbital period for IPs is, in general,
longer than that for polars with orbital periods above 3~h, where larger primary Roche lobes and higher mass-transfer rates (dM/dt) are expected. In the top and middle panels of Figure~\ref{magnetic}, we also include the period distribution for all IP and polar CVs with known orbital periods, shown in green \citep[see the 2026 version of these][]{MukaiIPCatalog,2025A&A...698A.106S}. Note that our sample follows the same trend as the complete sample.

Of the 135 previously classified as IP CVs in our sample, with a median orbital period of 14\,582~s (4.05~h), we have spin-period determinations for 82 systems. In Table~\ref{pspin}, we present the list of IP CVs with known spin periods, along with the value of P$_{\rm spin}$ and the orbital period. We determined the spin period for  82 IP CVS in this work. Of the 135 total, 83 objects have had their spin periods reported in the literature, specifically by \citet{1996ASSL..208..143H},  \citet{1983ApJ...264L..61P}, and \citet{2021MNRAS.507.6132P} (see column 5 in Table~\ref{pspin}). As expected for IP systems, the spin period is much shorter than the orbital period (10\% to 1\%) for most objects. Note that for 62 objects, we have confirmed the spin-period values reported in the literature within a 10\% uncertainty margin. 
The spin period distribution is included in the period distribution of IPs, shown in red in the top panel of Figure~\ref{magnetic}.  

Finally, the bottom panel of Figure~\ref{magnetic} shows the period distribution for single magnetic white dwarfs (Amorim in preparation). In this case, the period distribution exhibits a prominent peak at periods below 1 hour, consistent with the spin periods of IP CVs.  If we consider that mass transfer in CVs transports angular momentum towards the white dwarf, we would expect its rotation period to be shorter than that of single white dwarfs. However, the results show that for IPs, white-dwarf spin-up is not significant relative to the single-star case, as spin-down may also occur depending on the magnetosphere size relative to the disk corotation radius.
For polars, it is assumed that the magnetic field is strong enough to synchronize rotation with the orbital period. 
In the presence of strong orbital-synchronization torques, it is difficult to isolate the final effect of accretion on the white dwarf's spin.

%Magnetic braking plays a crucial role in the angular momentum loss (AML) mechanism driving CV evolution. The process occurs when ionized particles in the stellar wind follow magnetic ﬁeld lines, eﬀectively creating a lever arm that extracts angular momentum from the rotating star (Matt & Pudritz 2008). When MB ceases and AML decreases, the donor, which had previously expanded beyond its thermal-equilibrium radius, thermally relaxes and contracts within its Roche lobe. This temporarily halts mass transfer until gravitational radiation drives the system’s components back into contact with each other, allowing accretion to resume (Howell et al. 2001). The eﬃciency of this MB mechanism fundamentally determines the evolutionary pathways of CVs, including the location and width of the period gap (Knigge et al. 2011).

\section{AM CVns}

AM Canum Venaticorum stars (AM CVn) are binary systems with short orbital periods ranging from $\simeq$\,5 to 65 min. These systems are composed of white dwarfs accreting material from companion stars, typically a low-mass white dwarf or a semi-degenerate helium-rich star \citep[see, for instance,][]{Nather81,2013MNRAS.429.2143C,Ramsay18,Marcano21}.
We found 21 AM~CVn systems in our sample,
and the TESS photometric periods we detected are listed in Table~\ref{amcvn}.

\begin{table}
\scriptsize
    \centering
    \begin{tabular}{l|c|c|c|l|}
    \hline\hline
    Name& TIC& $P_\mathrm{TESS}$ (s)   &$P_\mathrm{orb}^\mathrm{lit}$ (s) &Reference\\
\hline
V803~Cen                 & 0030988774 & $1611.330\pm 0.007$   & $1596.2\pm 1.2$           &\cite{Roelofs07,Sun26} TESS superhump\\
GP Com                   & 0157201137 & $2794.43\pm 0.03$     & $2794.06\pm 0.17$         &\cite{Marsh99,Kupfer16}\\
AM CVn                   & 0321453404 & $1028.732\pm 0.004$+$1051.206\pm 0.019$ superhump& $1028.48\pm 0.14$  &\cite{Smak17,Nelemans02,Solheim98}\\ 
WDJ~120338.69-602247.98  & 0378898110 & superhump $1384.2831\pm 0.0014$+$1320.49\pm 0.04$ & 1320--1321 &\cite{Green24}\\
HP~Lib                   & 0386958844 & superhump $1118.80\pm 0.18$+$1102.7\pm1.3$ & $1102.70\pm 0.05$ &\cite{Patterson02} \\
ES~Cet                   & 0630055847 & $620.21528\pm 0.00012$& $620.21125\pm 0.00017$    &\cite{deMiguel18}\\
SDSS~J080449.49+161624.8 & 0800771427 & $10883\pm 4$          & $2670\pm 6$               &\cite{Roelofs09,Maccarone24}\\
ASASSN-19ct              & 0941094762 & $1857.588\pm 0.003$   & 1872                      &\cite{Green25}\\
SDSS~J131954.50+591514.6 & 0950467234 & $1268.49\pm 0.09$     & 3940                      &\cite{Ramsay18}\\
V396 Hya                 & 0951891812 & $64520\pm 5$          & 3906.0                    & \cite{Bravo25,Green25}\\
Gaiaaae                  & 1201247611 & $2982.470+/-0.004$ eclipse & 2982.6               & \cite{Campbell15,Maccarone24} \\
MGAB-V248                & 1883998117 & 1677.03389            & 1677.34                   & \cite{RiveraSandoval19,Dag26}\\
KL~Dra                   & 1884076062 & $1529.700\pm 0.003$   & 1500                      &\cite{Wood02}\\
\hline
OX Eri                   & 0650854785 & no detected period    & 2402.6                    &\cite{Green25}\\
YZ LMi                   & 0840677121 & no detected period    & $1698.7339778\pm 0.0000002$  &\cite{Schlindwein18}\\
V406 Hya                 & 0837081811 & no detected period    & 2028                      &\citep{Marcano21}\\
SDSSJ113732.34+405458.0  & 0900157782 & no detected period    & 3576.96                   &\cite{2023MNRAS.524.4867I}\\
SDSSJ120841.96+355025.2  & 0954326696 & no detected period    & $3180\pm 26$              &\cite{Green25}\\
SDSS~J141118.31+481257.6 & 1001089657 & no detected period    & $2765\pm 86$              &\cite{Anderson05}\\
SDSSJ150551.61+065948.5  & 1100144189 & no detected period    & $4061\pm 173$             &\cite{Green25}\\
ZTF18aavetqn             & 1716952687 & no detected period    & 1956                      &\cite{vanRoestel21}\\
\hline
 \end{tabular}
    \caption{AM~ CVn stars observed with TESS.}
    \label{amcvn}

\end{table}

\subsection{Orbital period distribution from the literature}
\label{sec-comparison}

\begin{figure}
    \includegraphics[width=0.76\textwidth,angle=270]{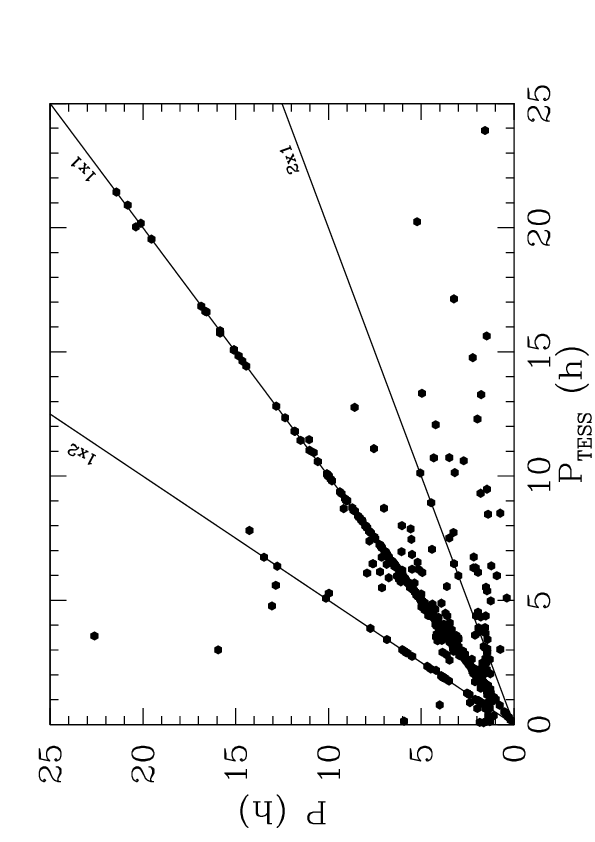}
\caption{Comparison between the periods determined in this work from photometric data and the 793 orbital periods reported in the literature. Only periods up to 25~h are considered in the plot (679 systems). The dashed lines indicate the $1\times 2$ (period from the literature is 2$\times$ the period from this work), $1\times 1$, and $2\times 1$ (period from this work is 2$\times$ the period from the literature) correspondences. The literature values for the orbital periods are taken mainly from \citet{2023AJ....165..163C,2023AJ....165..148H,2023MNRAS.524.4867I,2025A&A...698A.321W} (see the text for details).
    \label{comparison}}
\end{figure}

Of the 1362 CVs in our sample reported in this work, 793 also have orbital period determinations in the literature. Thus, we can compare our determinations of the periods obtained from the TESS photometric data with the orbital periods reported by other authors. In Figure~\ref{comparison}, we compare the periods reported in this work with those reported in other studies. For most objects, the values are taken from \citet{2023AJ....165..163C,2023AJ....165..148H, 2023MNRAS.524.4867I,2025A&A...698A.321W}, but we also use determinations from the works of \citet{1977ApJ...212L.125T,1982ApJ...254..646R,1986ApJ...305..732R,1996ASSL..208..143H,1998BaltA...7..287M,1998A&A...331..187M,2004PASP..116..300T,2010PASP..122.1285T,2017RNAAS...1...29T,2020AJ....160..151T,2001MNRAS.326..621N,2002AAS...201.4004B,2006A&A...449L..39N,2007RMxAA..43..291E,2008PASP..120..301B,2008JBAA..118..343S,2012Ap.....55..494P,2010PASJ...62..187U,2015MNRAS.452.1060C,2017NewA...50..109B,2017MNRAS.469.3688D,2022AJ....163..221J,2023ApJ...950..139G,2023A&A...678A.131K,2023MNRAS.523.3192M,2023ApJ...954..135W,2023MNRAS.519..352B,2024AJ....168..121B,2024MNRAS.535.3035D,2024A&A...687A.305M,2024ApJ...972...33S,Schaefer22, 2025arXiv250604371R}, indicated as ``other'' in Figure~\ref{comparison}. We show periods up to 25 hours since only 4\% of CVs have periods longer than this. As shown in Figure~\ref{comparison}, for most objects, the periods determined in this work are consistent with the values reported in the literature and lie along the 1$\times$1 correspondence line. In fact, 72\% have orbital periods consistent with those reported in previous work, within 30\%. For 50 objects, the period determined in this work is half ($\pm 10\%$) of that from the literature, and for 6 objects, the period is twice ($\pm 20\%$), corresponding to the 1$\times$2 and $2\times$1 lines in Figure~\ref{comparison}, respectively.

\begin{figure}
    \includegraphics[width=0.97\textwidth]{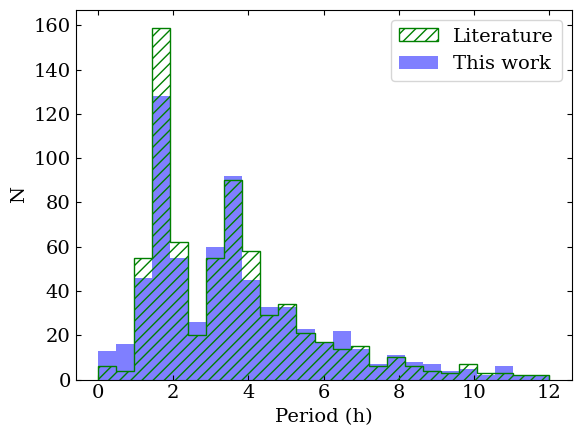}
 \caption{Distribution of the orbital period for CVs, with periods determined in the literature (green) {\bf and the photometric ones} in our work (blue). We show only systems with periods up to 12~h (see the text and Figure~\ref{comparison} for details).}
   \label{comparison-hist}
\end{figure}

Figure~\ref{comparison-hist} compares the orbital period distributions from this work (green) with those from the literature (blue) for the 793 CVs in our sample with previous orbital period determinations. Both distributions are similar, showing the expected gap between $\sim$2 and $\sim$3~h. In the literature distribution, there is also a larger contribution of CVs with orbital periods below the gap, between $\sim$1 and $\sim$2~h, compared to values just above the period gap. This difference is reduced for the distribution obtained in this work. This effect alters the median values of the distributions: 13\,226~s (3.674~h) for the period distribution in this work and 12\,428~s (3.452~h) for the values reported in the literature. We also found an increase in periods below 1~h compared to previous determinations, thereby increasing the sample of AM~CVn-type binaries.

\subsection{Comparison with non-interacting white dwarf stars}

For single stars, angular momentum loss via magnetic braking will shorten the star's rotation period \citep{1972ApJ...171..565S}. 
For binaries, the tidal forces prevent the non-degenerate component from rotating more slowly; thus, angular momentum is transferred from the orbit, causing the orbital period to decrease \cite[e.g.][an references therein]{Sun26}.
%Further tidal interactions make the secondary rotate faster. 
Hence, we can expect that single stars will rotate more slowly than those in binaries. The large uncertainties in the angular momentum transfer mechanisms and magnetic braking during the evolution of both single stars and binaries make comparison a tool in the study of angular momentum transfer \citep[e.g.][]{Tayno21,2024ApJ...974..314O,Zhou26}.

%\citet{1977A&A....57..383Z} analyzed the physical mechanisms that produce tidal friction in binaries and determined that the synchronization time increases rapidly with the orbital period. The author suggested that binary systems with orbital periods shorter than 17 days should be synchronized. More recent computations, which consider binary systems composed of compact objects such as white dwarf stars, suggest a significantly shorter orbital period for synchronization, on the order of $\sim$1~h \citep{2012MNRAS.422.1343P, 2013MNRAS.430..274F}.

\begin{figure}
\includegraphics[width=0.95\textwidth]{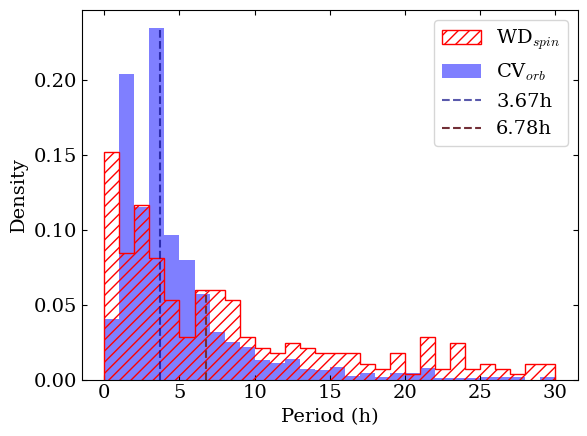}
\caption{Comparison of the period distributions for CVs (blue) and variability periods for white dwarf stars (red) from \citet{2024ApJ...974..314O} up to 30\,h. The dashed lines are the median of each distribution.
    \label{rotation}}
\end{figure}

\citet{2024ApJ...974..314O} presented a study of the variability periods of a population of 318 white dwarf stars using TESS photometric data. This sample contained single, likely single, and binary systems with variability periods ranging from 0.13~h to 270~h. The authors propose that the observed variability periods are rotation periods, but they cannot rule out orbital or synchronized periods among the systems. In Figure~\ref{rotation} we show the period distribution for the sample presented in \citet{2024ApJ...974..314O}  (red), along with the orbital period distribution for the 1362 CVs analyzed in this study (blue), with periods up to 30~h. 

The median value for the complete sample from \citet{2024ApJ...974..314O} is 6.803~h, which is twice the median orbital period of the CVs in our sample (13\,226\,s or 3.674~h). 
In addition, for WDs with pairs, the median variability period is $\sim$7.8~h, which is twice the median orbital period of eclipsing CVs. 

For the 115 classified as likely single WDs, \citet{2024ApJ...974..314O} reported a median of 3.9 hours, which is closer to the CV median. This result is similar to what we find for magnetic white dwarfs in Section~\ref{sec3.5}, where single magnetic white dwarfs show rotation periods similar to the spin periods of IP CVs.

Finally, for magnetic systems, \citet{2024ApJ...974..314O} reported a median variability period of 3.4~h. In contrast, for the magnetic CVs in our sample, the values were 1.91~h and 4.05~h for polars and IPs, respectively.

\section{Summary}
\label{summary}

We determined photometric periods with a false alarm probability of less than 1/1000 for 1362 known CVs, including 564 for which periods had not previously been determined. Our determinations were obtained from 120\,s and 20\,s cadence photometry with the TESS Space Telescope, up to sector 101. To our knowledge, this is the largest number of CVs with homogeneously determined periods.

We detected photometric periods from 253.6\,s to 258.6\,h, with $\sim$92\% of the sample showing orbital periods below 12~h.
From the photometric period distribution of the 1362 CVs in our sample, we recover the well-known ``period gap'' between 2 and 3~h, where the number of CVs drops sharply due to the halt in mass transfer. The median period is 3.689~h. Our sample includes CVs that exhibit one or two eclipses in their light curves, providing a robust determination of the orbital period. Most periods are between $\sim$2 and 15~h. The period distribution for eclipsing systems also shows a well-defined period gap, reproducing the result obtained for the complete sample. 
%Finally, we present 25 objects that could be classified as CVs based on the shape of their light curves; however, spectroscopic confirmation is needed. 

For magnetic CVs, polars, and IPs, a period gap is not expected due to a strong magnetic field that 
controls the incoming material from the secondary star, 
fully or partially suppressing the formation of an accretion disk
\cite[e.g.][]{Zhou26}. We determine the spin periods for 82 IPs, including 17 new determinations. As expected, the spin period is 0.1---0.01\% of the orbital period in all cases. 
For the 62 polars in our sample, the orbital periods are shorter than 7.5~h, with a median value of 1.91~h. On the other hand, for the 135 IPs analyzed in this work, the orbital periods range up to $\sim$44.774~h, with a median of 4.05~h. No clear evidence of a period gap was observed in our sample of classical novae. However, short-period nova systems are dominant.

We compare the distribution of rotation periods for single magnetic white dwarfs with those of the polar and IP CVs in our sample. For polars, we assume the magnetic field is strong enough to synchronize the white dwarf's rotation period with the orbital period.
%In contrast, for IPs, the rotation period of the white dwarf component is determined by its spin period. 
We find that the period distribution of single magnetic white dwarfs peaks at 1~h, with values similar to their spin periods, implying that the spin-up mechanisms for both are equally efficient.

For the 793 CVs with prior orbital-period determinations, we compare our results with those in the literature. For $\sim$ 72\% of the objects, our determinations match those from previous works within 30\%. The period distributions are similar in shape and are characterized by a defined period gap. However, the period distribution in this work shows a gap that is populated by a factor of $\sim 2$ compared to the literature. 

Finally, we found that the median photometric period in our CV sample is half of the median of the rotation distribution of white dwarfs reported by \citet{2024ApJ...974..314O}. Also, for magnetic single white dwarfs, we found that the rotation period distribution is similar to that of the spin period for IP CVs, with a large contribution of periods below $\sim$~1~h.

\section*{Acknowledgements}
This work was financed by CNPq-Brazil, CAPES-Brazil, and FAPESP-Brazil.
This paper includes data collected with the TESS mission, obtained from the MAST data archive at the Space Telescope Science Institute (STScI). Funding for the TESS mission is provided by NASA's Science Mission Directorate. STScI is operated by the Association of Universities for Research in Astronomy, Inc., under NASA contract NAS 5–26555. 
This research has made extensive use of the Astrophysics Data System, funded by NASA under Cooperative Agreement 80NSSC21M0056.
ADS was used extensively during this work. This work includes results from the European Space Agency (ESA) space mission Gaia. Gaia data are being processed by the Gaia Data Processing and Analysis Consortium (DPAC). Funding for the DPAC is provided by national institutions, particularly the institutions participating in the Gaia Multilateral Agreement (MLA). The Gaia mission website is https://www.cosmos.esa.int/gaia. The Gaia archive website is https://archives.esac.esa.int/gaia. This work made use of Astropy: \footnote{https://www.astropy.org}, a community-developed core Python package and an ecosystem of tools and resources for astronomy \citep{astropy:2013, astropy:2018, astropy:2022}. We used the TESS-LS software \footnote{\url{https://github.com/ipelisoli/TESS-LS}}
to retrieve the TESS light curves and perform the period search, and we used TESS-Localize \citep{Higgins2022} to identify the source of the observed variability. 
Most objects were included in TESS proposals G011268,
G022071,
G03071,
G04046,
G05094,
G06027,
G07025,
G08036, by PI: Simone Scaringi (University of Durham).

\section*{Data Availability Statement} 

All data analysed here are available in a public archive, the MAST data archive at the Space Telescope Science Institute (STScI).
We provide the data described here in Supplementary files: kepler26.tsv and kepler.n.

\bibliographystyle{mnras}
\bibliography{cv-tess}{}

\appendix

\onecolumn
\section{List of IP CVs and suspected IPs}
\label{AppendixA}
\LTcapwidth=\textwidth
\begin{longtable}{ccrrrr}
\caption{List of IP CVs and suspected IPs, with measured periods from this work and the literature. The TIC is shown in column 1, the name in column 2, the orbital period determined in this work is listed in column 3, column 4 shows the values for the spin period from this work, while columns 5 and 6 show the values taken from the literature \protect\citep{1983ApJ...264L..61P, 1996ASSL..208..143H, 2021MNRAS.507.6132P, Suleimanov25}. All periods are in seconds.}
\label{pspin}\\

\hline
TIC & Name & $P_{\rm orb}$ & $P_{\rm spin}$ &
$P_{\rm orb}^{\rm lit}$ & $P_{\rm spin}^{\rm lit}$ \\
\hline
\endfirsthead

\multicolumn{6}{c}%
{{\tablename\ \thetable{} -- continued from previous page}} \\

\hline
TIC & Name & $P_{\rm orb}$ & $P_{\rm spin}$ &
$P_{\rm orb}^{\rm lit}$ & $P_{\rm spin}^{\rm lit}$ \\
\hline
\endhead

\hline
\multicolumn{6}{r}{{Continued on next page}} \\
\endfoot

\hline
\endlastfoot
0003645653 & ATCnc                   & $(17434.12\pm 0.80)$~s   & $(1415.70\pm 0.03)$~s?      & 17375.0s    & 1604s \\
0005657040 & HD58756                 & $(38261\pm 18)$~s        &                             & 15529.3s    & \\
0009464138 & V795Her                 & $(9353.83\pm 0.11)$~s    &                             & 9354.1s     &\\
0009560142 & EXHya                   & $(5894.868\pm 0.017)$~s  & $(4021.573\pm 0.002)$~s     & 5904.0s     & 4021.6162s\\
0011116617 & RXJ2015.6+3711          & $(22937.30\pm 0.16)$~s   & $(7198.40\pm 0.08)$~s       & 45940.0s    & 7196s \\
0015556256 & V2467Cyg                & $(13282\pm 9)$~s         &                             & 13752s?     & 2078.4s? \\
0017745208 & PQGem                   & $(18694.16\pm 0.26)$~s   & $(833.488\pm 0.010)$~s      & 18693.4s    & 833.41992s \\
0018053590 & V426Oph                 & $(24695\pm 42)$~s        &                             & 24649.9s    & \\
0019028616 & DWCnc                   & $(5103.462\pm 0.027)$~s  & $(2315.0130\pm0.002)$~s     & 5165.856s   & \\
0021162863 & XSSJ12270-4859          & $(24873.530\pm 0.054)$s  &                             & 24866.96s   & \\
0021505340 & TXCol                   & $(20463.0\pm 1.1)$s      & $(1909.3531\pm 0.046)$~s    & 20589s      & 1909.5s \\
0022641187 & V348Pup                 & $(8798.889\pm 0.004)$s   &                             & 8798.89s    & \\
0024450633 & TVCol                   & $(19756.08\pm 0.04)$s    &                             & 19751.04s   & 1909.7s\\
0025611385 & SwiftJ0939.7-3224       & $(30611\pm 41)$s         & $(2671.55\pm 0.46)$s        & 2669.76s    & 2670s\\
0035975843 & SwiftJ0958.0-4208       & $(21974\pm 23)$s         & $(296.122\pm 0.028)$s       &             & 295.22s \\
0037430543 & VZPyx                   & $(6531.152\pm 0.009)$s   & $(303.770\pm 0.001)$s?      & 6334.9s     &\\
0038305527 & FOAqr                   & $(17455\pm 8)$s          & $(1254.379\pm 0.053)$s      & 17458.0s    & 1254.330973s\\
0053180626 & V405Aur                 & $(14907.870\pm 0.24)$s   & $(545.480\pm 0.002)$s       & 14914.368s  & 545.4555s\\
0055381011 & HS0922+1333             & $(14541.85\pm 0.26)$s    &                             & 14542.2s    &\\
0059627386 & HYLeo                   & $(25396\pm 85)$s         & $(4021.78\pm 4.74)$s        & 15911.4s    & 4150s?\\
0062698629 & SwiftJ2116.5+5336       & $(23599.94\pm 0.04)$s    &                             & 23599.9s    &\\
0062845887 & YYSex                   & $(5666.799\pm 0.001)$s   &                             & 5556.293s   & 1444s \\
0065820714 & IGRJ04571+4527          & $(21954.72\pm 0.91)$s    & $(1218.695\pm 0.001)$s      & 22275.6s    & 1218.7s\\
0067482244 & SDSSJ231909.18+331539.6 & $(6312.18\pm 0.05)$s     &                             & 12628.224s  & \\
0068722402 & IGRJ19267+1325          & $(12478\pm 5)$s          & $(935.31\pm 0.02)$          & 12415.7s    &\\
0071013469 & AHEri                   & $(10360\pm 16)$s         &                             & 20658.2s    &\\
0073319029 & V381Vel                 & $(8051.453\pm 0.003)$s   &                             & 8043.84s    & 7388s?\\
0073535720 & RXJ1404.4+1723          & $(5690\pm 4$)s           &                             &             &\\
0075654576 & V1323Her                & $(17450.51\pm 0.14)$s    & $(1520.3817\pm 0.0001)$s    & 15845.8s    & 1520.51s\\
0076191504 & V533Her                 & $(12733.71\pm 0.06)$s    &                             & 12700.8s    & 63.633s?\\
0076381942 & FSAur                   & $(12326.54\pm 0.04)$s    &                             & 5148s       &\\
0077841332 & UUCol                   & $(12469\pm 10)$s         & $(863.77\pm 0.06)$s         & 12420s      & 863.5s \\
0079650089 & 1RXSJ080114.6-462324    & $(42491.4\pm 1.3)$s      & $(1307.4893\pm 0.0004)$s    & 42490.8s    & 1307.517s \\
0086408822 & WZSge                   & $(4897.84\pm 0.032)$     & $(699.6689\pm 0.0008)$s     & 4897.8s     &\\
0091734151 & V418Gem                 & $(15728.87\pm 0.20)$s    & $(480.6693\pm 0.0001)$s     & 15733.4s    & 480.6700s\\
0092440509 & 1RXSJ205652.1-301433    & $(6363.45\pm 0.11)$s     &                             & 6363.619s   & 29.6s\\
0093196973 & IGRJ08390-4833          & $(21955.77\pm 0.03)$s    & $(1589.5990\pm 0.0002)$s    & 28512.0s    & 1480.8s\\
0097328148 & TTAri                   & $(11485.05\pm 0.14)$s    &                             & 11884.3s    &\\
0102731925 & V2306Cyg                & $(15737.47\pm 0.06)$s    & $(733.34250\pm 0.00009)$s   & 15737.5s    & 1466.67952s\\
0103605282 & SDSSJ100516.61+694136.5 & $(13134.59\pm 0.22)$s    &                             & 13115.52s   &\\
0117885256 & WXPyx                   & $(26831.99\pm 0.31)$s    & $(1557.2794\pm 0.0002)$s    & 19932.5s    & 1559.2s \\
0122645363 & IGRJ17014-436           & $(46140.44\pm 0.19)$s    & $(1858.7280\pm 0.0006)$s    & 46141.1s    & 1859.11s\\
%\end{tabular}
%
%	\label{pspin}
%\end{center}
%\end{deluxetable*}
%
%\begin{deluxetable*}{ccccc}
%\caption{continuation of Table~\{pspin}}
%	\begin{tabular}{cccccc} % four columns, alignment for each
%\hline
% TIC       &  Name &  P$_{\rm O}$   & P$_{\rm spin}$ &  P$_{\rm O}^l$   & P$_{\rm spin}^l$ \\\\
% \hline
%
0136171236 & IGRJ21095+4322          &                          &  $(459.380\pm 0.001)$s?     &             & \\
0138042556 & LSCam                   & $(11865.59\pm 0.03)$s    &                             & 12302.0892s & \\
0139144802 & V2069Cyg                & $(26930.73\pm 0.34)$s    & $(743.369\pm 0.002)$s       & 26929.2s    & 743.40650s \\
0142155545 & EIUMa                   & $(23173.9\pm 1.6)$s      & $(769.72\pm 0.05)$s         & 23148s      & 741.6s\\
0142864870 & DODra                   & $(14288.50\pm 0.11)$s    & $(529.330\pm 0.002)$s       & 14288.3s    & 529.31s\\
0144324239 & V1025Cen                & $(5077.9\pm 1.1)$s       & $(2146.69\pm 0.03)$s        & 5076.9s     & 2146.59s \\
0145620527 & CPPup                   & $(5304.33\pm 0.01)$s     &                             & 5292.0s     &\\
0147178211 & 1RXSJ065806.3-174427    & $(8569.25\pm 0.01)$s     &                             & 8565.3s     &\\
0153352161 & HZPup                   & $(18448\pm 12)$s         & $(1211.8\pm 0.3)$s          & 18324s      &\\
0154485903 & [DWS97]Dra7             & $(43314\pm 13)$s         & $(3856.65\pm 0.02)$s        &             &\\
0157260715 & SWUMa                   & $(4920.77\pm 0.03)$s     & $(1048.544\pm 0.006)$s      & 4908.8s     &\\
0165952825 & Lanning386              & $(14174.21\pm 0.03)$s    & $(833.770\pm 0.001)$s       & 14174.1s    & \\
0172503328 & V647Aur                 & $(12476.94\pm 0.11)$s    & $(932.9019\pm 0.0005)$s     & 12479.616s  & 932.9123s \\
0173460141 & V603Aql                 & $(11982\pm 20)$s         &                             & 11940.6s    &\\
0191430151 & IGRJ16500-3307          & $(13015.81\pm 0.06)$s    & $(571.9003\pm 0.0009)$s     & 13020.5s    & 571.9s\\
0193515431 & V1460Her                & $(17958.56\pm 0.10)$s    &                             & 17958.4s    & 38.875s\\
0196278926 & V515And                 & $(9829.3\pm 0.2)$s       & $(465.4695\pm 0.0005)$s     & 9831.9s     & 465.5s\\
0199296334 & HBHA4705-03             & $(14844.48\pm 0.02$)s    &                             & 14844.946s  &\\
0218971180 & CScl                    & $(5059.88\pm0.01)$s      & $(389.4718\pm 0.0005)$s     & 5060.2089s  &\\
0221227893 & IGRJ16547-1916          & $(13370\pm 5)$s          & $(546.65\pm 0.02)$s         & 13374.7s    & 546.6606s \\
0224222776 & AOPsc                   & $(12927.67\pm 0.11)$s    & $(805.128\pm 0.007)$s       & 12927.7s    & 805.2024s\\
0224325028 & V1084Her                & $(10105.3\pm 0.2)$s      &                             & 10416.4s    & 1903s? \\
0234386436 & BGCMi                   & $(11641.37\pm 0.02$)s    & $(913.4618\pm 0.0001)$s     & 11642.2s    & 913.5s\\
0234712743 & V902Mon                 & $(29376.34\pm 0.04$)s    & $(2207.740\pm 0.001)$s      & 29383.2s    & 2210s\\
0236868718 & SDSSJ201116.83+600428.0 & $(8839.32\pm 0.02)$s     & $(3637.294\pm 0.001)$s      & 8838s       & 3637.2s\\
0248138854 & AEAqr                   & $(35679\pm 50$)s         &                             & 35567.1s    & 33.08s \\
0261376909 & AHMenA,B                & $(10662.11\pm 0.04)$s    &                             & 10620.0s    & \\
0261698173 & IGRJ15094-6649          & $(21146.74\pm 0.12)$s    & $(809.50971\pm 0.00009)$s   & 21202.6s    & 809.42s\\
0268294933 & V436Car                 & $(15748.3\pm 0.3)$s      &                             & 12960s      &\\
0271137877 & SDSSJ075653.11+085831.8 & $(11834.50\pm 0.03)$s    &                             & 11852.4     &\\
0277479500 & V1223Sgr                & $(12115\pm 4)$s          & $(745.5\pm 0.2)$s           & 12117.1s    & 745.5.6s\\
0281412181 & CWMon                   & $(16718\pm 40)$s         &                             & 15258.24s   &\\
0289786644 & V2275Cyg                & $(39980\pm 3)$s          & $(1476.478\pm 0.007)$s      & 27171.936s  & 1475s?\\
0302582144 & V2400Oph                & $(13692\pm 13)$s         & $(927.72\pm 0.05)$s         & 12348s      & 927.66s\\
0304299578 & PBCJ0927.8-6945         & $(17350.94\pm0.05)$s     & $(1032.9737\pm 0.0004)$s    & 17461.4s    & 1033.05s\\
0307442311 & V552Aur                 & $(34102\pm 6)$s?         &                             & 5259.0s     &\\
0311987673 & GZCnc                   & $(8030.782\pm 0.10)$s    & $(360.470\pm 0.012)$s       & 7611.84s    &\\
0313242194 & MUCam                   & $(16989.60\pm 0.04)$s    & $(1187.2005\pm 0.0003)$s    & 17008.1856s & 1187.2184s\\
0314714161 & 1RXSJ230645.0+550816    & $(27736\pm 46)$s?        & $(464.458\pm0.005)$s        & 11750.4s    & 464.452s\\
0316276691 & BZUMa                   & $(5861.90\pm 0.08)$s     &                             & 5874.3s     &\\
0320180973 & V709Cas                 & $(19198.5\pm 0.4)$s      & $(312.7500\pm 0.0003)$s     & 19198.4s    & 312.7478s\\
0332624864 & DQHer                   & $(16728.81\pm 0.06)$s    & $(71.0600\pm 0.0004)$s      & 16740.0s    & 71.065583s\\
0333323257 & V455And                 & $(4823.187\pm 0.009)$s   & $(67.27000\pm 0.00006$)s    & 4865.1s     & 67.619685s\\
0334061567 & SwiftJ0746.2-1611       & $(33716\pm 4)$s          & $(2248.855\pm 0.006)$s      & 33783.3s    & 2249.0s\\
0335654642 & HTCam                   & $(5158.43\pm 0.04)$s     & $(515.0595\pm 0.0002)$s     & 5159.1s     & 515.0592s\\
0349069732 & V1062Tau                & $(3593.5\pm 1.0)$s       & $(3684.734\pm 0.009)$s      & 35936.0064s & 3780s \\
0350297825 & TWPic                   & $(25040.5\pm 0.2)$s      & $(8048.07\pm 0.08)$s        & 21816s      & 7186s? \\
0367086580 & SwiftJ0525.6+2416       & $(39289\pm 39)$s         &                             &             & 226.3s \\
0380576494 & RZLeo                   & $(6569.0\pm 0.6)$s       &                             & 6569.7091s  & \\
%0382408105 & 1ES1210-646             & 595186.75s & 12696.63s? & 578707.2s   &\\ %(Masetti et al.; 2009, A&A, 495, 121; 2010, A&A,519, A96
%\end{tabular}
%\end{deluxtable*}
%
%\begin{deluxetable*}{ccccc}
%\caption{continuation of Table~\{pspin}}
%	\begin{tabular}{cccccc} % four columns, alignment for each
%\hline
 %TIC       &  Name &  P$_{\rm O}$   & P$_{\rm spin}$ &  P$_{\rm O}^l$   & P$_{\rm spin}^l$ \\\\
 %\hline
0383534701 & IGRJ14257-6117          & $(14490\pm 4)$s           &                            & 14601.6s    & 509.5s\\
0385080312 & LSPeg                   & $(14623.4\pm 0.3)$s       &                            & 12960s      & \\
0387201824 & 2MASSJ09213414-5939068  & $(10930.23\pm 0.04)$s     & $(907.97910\pm 0.00009)$s  & 10946.9s    & 908.12s \\
0387448087 & SGRAJ204547.8+672642    & $(10729.10\pm 0.18)$s     & $(844.91\pm 0.03)$s?       & 10720s?     &\\
0395587587 & RXJ0153.3+7446          & $(14169.5\pm 0.7)$s       &                            & 14182.56s   & 1974s? \\
0396998901 & QZVir                   & $(5696\pm 8)$s            & $(2390.9\pm 1.6)$s?        & 5082.1s     &\\
0397773193 & V1082Sgr                & $(75411\pm 203)$s         & $(6400\pm 12)$s            & 74953.7s    & 6995s\\
0403308597 & V592Cas                 & $(10645.13\pm 0.15)$s     &                            & 9936s       &\\
0406964858 & V1033Cas                & $(14510.2\pm -.4)$s       & $(563.1200\pm 0.0003)$s    & 14523.84s   & 563.5s \\
0410152621 & SSS110720:224200-662512 & $(35150\pm 34)$s          & $(2675.1\pm 0.2)$s         &             &\\
0411357254 & IGRJ12123-5802          & $(23295.1\pm 1.1)$s       &                            & 27388.8s    &\\
0411381210 & IGRJ16167-4957          & $(17997\pm 23)$s          & $(582.46\pm 0.2)$s         & 18014.4s    & 582.45s \\
0411603422 & V842Cen                 & $(12798.4\pm 0.5)$s       &                            & 14169.6s    & 56.825s\\
0415937547 & IGRJ14536-5522          & $(11361.790\pm 0.010)$s   &                            & 11363.1s    &\\
0417434471 & RXJ2133.7+5107          & $(24235\pm 3)$s           & $(570.80\pm 0.02)$s        & 25698.04s   & 570.8189s\\
0431762266 & GKPer                   & $(171932\pm 10)$s         & $(351.320\pm 0.002)$s      & 172454.0s   & 351.332s\\
0434329100 & SwiftJ0614.0+1709       & $(18336.59\pm 0.09)$s     & $(1411.6522\pm 0.0006)$s   &             & 1412.3s\\
0451182728 & APCru                   & $(18675.3\pm 0.4)$s       &                            & 18403.2s    & 1837s \\
0455269830 & VZSex                   & $(12877.2\pm 0.5)$s       & $(2619.49\pm 0.05)$s       & 12864.96s   & 2450s?\\
0686160823 & SwiftJ0503.7-2819       & $(4897.963\pm 0.006)$s    & $(3932.249\pm 0.010)$s     & 4897.66596s & 3932.03555s \\
0734907771 & TXCol                   & $(20494.1\pm 0.3)$s       & $(1909.544\pm 0.003)$s     & 20589.1s    & 1909.5s\\
0750058684 & V667Pup                 & $(20142\pm 39)$s          & $(512.39\pm 0.04)$s        & 20200.3s    & 512.4s\\
0801282759 & SDSSJ084617.12+245344.1 & $(15790\pm 8)$s           &                            & 16675.2s    &\\
0867320307 & KOVel                   & $(36334\pm 39)$s          & $(5568.2\pm 0.7)$s         & 36296.64s   &\\
0938248761 & [PK2008]HalphaJ115927   & $(25953\pm 41)$s          & $(1161.48\pm 0.06)$s       & 25920s      & 1215.99s \\
1003231665 & RRCha                   & $(12113.72\pm 0.03)$s     & $(1728.988\pm 0.002)$s     & 12104.64s   & 1950s \\
1016622332 & V1039Cen                &                           & $(501.517\pm 0.013)$s?     & 21340.8s    & 720s?\\
1027495091 & IGRJ14091-6108          & $(57999\pm 489)$s         & $(576.64\pm 0.06)$s        & 57012.768s  & 576.623s\\
1201319601 & ZTF18abaaewz            & $(5585.62\pm 0.08)$s      & $(3392.87\pm 0.04)$s       & 5585.65s    & 3392.32s\\
1610654644 & IGRJ18173-2509          & $(5517.11\pm 0.8)$s       &                            &             & 1663.4s\\
1658661399 & IGRJ18308-1232          & $(19438\pm 14)$s          &                            & 19431.36s   & 1820s\\
1779900865 & AXJ1853.3-0128          & $(19373\pm 22)$s?         &                            & 5235s       & 476.0s \\
1798237213 & PBCJ1911.4+1412         & $(23365\pm 9)$            & $(746.8506\pm 0.0016)$s    & 23358.24s   & 746.885s\\
1976538531 & 1RXSJ211336.1+542226    & $(13124.81\pm 0.02)$s     & $(1267.5821\pm 0.0006)$s   & 15000s      & 1265.4s \\
2017894254 & IPHASJ213849.91+554405.6& $(15658.08\pm 0.09)$s     & $(989.1356\pm 0.0005)$s    & 15923.5s    & 989.43s \\
2024398006 & V349Aqr                 & $(11618.37\pm 0.18)$s     &                            & 11658.1248s & 390.15s\\
2053498035 & V598Peg                 & $(5432\pm 2)$s            & $(1263.01\pm 0.03)$s       & 4985.28s    & 2499.6s
\end{longtable}

\begin{landscape}
\LTcapwidth=\textwidth
\scriptsize
\begin{longtable}{cccccl}
\caption{List of 1554 CVs with TESS 120~s data examined in this work. The TIC number, sectors observed, RA, DEC, FAP(1/1000) in mma (ppt), are listed in columns 1 to 5. The photometric periods are listed in column 6. Alternative names, literature periods and source, Gaia DR3 ID, G, par (mas), MG, bp-rp, comments on the characteristics of the CVs and CROWDSAP are in column 7. 
We do not attempt to characterize the mechanism underlying all observed periodicities. We only list those TESS\_Localize indicates that are coming from the CV. Only a portion of this table is shown here to demonstrate its form and content. A machine-readable version of the full table is available in file kepler26.tsv 
%doi:10.5281/zenodo.21208066 
(Kepler\_h for periods in hours).
\label{table-results}} \\
\hline
\multicolumn{6}{c}{\textbf{TIC, Sectors, RA, DEC, FAP, Periods}} \\
\multicolumn{6}{c}{\textbf{Name, Gaia DR3, Reference, CROWDSAP}} \\
\hline
\endfirsthead
\multicolumn{6}{c}{{\bfseries \tablename\ \thetable{} -- continued from previous page}} \\
\hline
\multicolumn{6}{c}{\textbf{TIC, Sectors, RA, DEC, FAP, Periods}} \\
\multicolumn{6}{c}{\textbf{Name, Gaia DR3, Reference, CROWDSAP}} \\
\hline
\endhead
\hline
\multicolumn{6}{r}{{Continued on next page}} \\
\endfoot
\hline
\endlastfoot

% DADOS DA TABELA (primeiras linhas como exemplo)
0003034524&19,43-45,59,71-72&82.141990&33.306049&10.01&30918.31s@158.19mma double eclipse,2f1=15443.87s@237.22mma+10harmonics\\
&&QZAur & 3449050362952844288  & Canbay23 P=30887.7s &  CROWDSAP:0.075 \\
0003399307&54,80&289.992625&-7.1819444&3.76&16598.43s@39.22mma,2f1=8485.88s@9.86mma\\
&&CZAql AN47.1925 &4207860450105410688  & Canbay23 P=17323.2s & CROWDSAP:0.221  \\
0003457576&59,73,86&082.707600&35.912400&7.13&9690.84s@12.66mma,22480.48s@20.15mma,151971.56s@97.59mma\\
&&IVAur 2MASSJ05304981+3554446 & 183305733863654144 & Canbay23 noper  & CROWDSAP:0.132 \\
0003547397&19,43-45,71-72&82.9964167&30.4458333&1.78&17658.25s@95.36mma,2f1=8829.16s@19.88mma+5harm\\
&&TAur HD36294 & 3446266197646225536 & Canbay23 P=17658.3s & CROWDSAP:0.402 \\
0003645653&21,44-47,f71-f72&127.1538333&25.3340556&0.860&19445.49s@91.24mma,2f1=9741.46s@4.39mma+3harm,17434.27s@6.08mma,\\
&&&&&Pspin=1415.63s@0.91mma,103569.21s@9.80mma,146349.18s@15.88mma,\\
&&&&&197347.15s@21.61mma,burst bjd=1850\\
&&ATCnc SDSSJ082836.92+252003.0 & 679528804789642240 & Canbay23 P=17375.0s Pspin=1600s Bruch19 IP & CROWDSAP:0.973 \\
0003801855&49&169.264700&+18.432820&1.48&18726.38s@2.70mma,37430.55s@2.66mma,152046.16s@44.41mma\\
&&SDSSJ111703.52+182558.1 HKLeo &
3971876285214587136 & Hou23 P=152057.347s & CROWDSAP:0.995 \\
0004896794&21,48&147.95425&34.1235&5.09&5150.06s@17.85mma,267.91s@9.01mma,186537.55s@117.00mma,\\
&&&&&2113488.04s@753.22mma\\
&&RZLMi PG0948+344 & 795244943252669568 & Canbay23 P=5045.8s & CROWDSAP:0.963 \\
0004970499&35-36,62,89&141.966487&-39.181266&55.68&9734.95s@11640.3mma,2f1=4867.48s@216.59mma,\\
&&&&&19469.95s@1345.08mma,328.52s@69.19mma,3893.98s@27.41mma,\\
&&&&&2433.64s@16.07mma\\
&&2MASSJ09275195-3910524 & 5429765829625368832 & Canbay23 P=14688s & CROWDSAP:0.036 \\
0005638374&58,85&32.3744167&28.5415556&14.66&5777.44s@221.77mma double complex eclipse,2f1=2888.99s@185.54mma,\\
&&&&&1444.36s@2.04mma+harmonics \\
&&BSTri 1RXSJ020928.9+283243 & 107522105369340800 & Canbay23 P=5778.52s & CROWDSAP:0.535 \\
0005657040&34,f80&111.6960417&-6.6749167&6.27&38635.49s@19.75mma,2f1=19355.85s@22.04mma \\
&& GIMon HD58756 & 3054845658802097280 & Canbay23 P=15569.3s IP & CROWDSAP:0.100\\
\end{longtable}
\end{landscape}

\newpage
\section*{APPENDIX B: CV STARS WITH NEWLY DETECTED PHOTOMETRIC PERIOD}
\addcontentsline{toc}{section}{APPENDIX B: CV STARS WITH NEWLY DETECTED PHOTOMETRIC PERIOD}
\label{AppendixB}
\begin{table}
\centering
\begin{tabular}{|l|l|r|}
\hline
\multicolumn{3}{|c|}{\textbf{TIC \hfill Name \hfill $P_\mathrm{main}$ (s)}} \\
\hline
0003457576 & IVAur & $9690.8\pm 0.3$\\
0006348255 & RSOph & $3269\pm 6$\\
0007551387 & GSC06742-00051 & $11990.71\pm 0.08$ eclipse\\
0011114823 & RXJ2015.2+3659 & $12499.7\pm 02$\\
0012524740 & WDJ032539.43-081442.77 & $7524.92\pm 0.11$\\
0014890698 & 1RXSJ165032.3-242643 & $23901\pm 32$\\
0015853131 & 2MASSJ04070399+1855380 & $ 10560.82\pm 0.08$\\
0018055392 & V0952Oph & $ 9720\pm 14 $\\
0019488490 & GALEXJ091803.6-303547 & $ 12683\pm 17$\\
0022021279 & ASASSN-14km & $ 15749.9\pm 0.3 $ eclipse\\
0025471934 & BVAnd & $ 10090.01\pm 0.17 $\\
0026429577 & ASASSN-19bp & $ 30235.2\pm 0.6 $\\
0026576539 & ASASSN-19zo & $ 22070\pm 51 $\\
0026623330 & CRTSJ035034.9+353247 & $ 30194\pm 97$\\
0027583971 & NSV1586 & $ 12528\pm 25$\\
0029063564 & MGAB-V279 & $ 11901.4\pm 0.5 $\\
0029211429 & ASASSN-17cw & $ 6813.91\pm 0.06 $\\
0029290702 & DDE170 & $ 3245.9\pm 0.6 $\\
0030927975 & ASASSN-18abu & $ 15773\pm 3$ eclipse\\
0033366214 & V838Mon & $ 56476\pm 255 $\\
0035975843 & SWIFTJ0958.0-4208 & $ 21974\pm 23 $\\
0035976533 & GSC07702-02511 & $ 27926.6\pm 0.4 $\\
0036434992 & Gaia19bhh & $ 25210\pm 33 $\\
0037176324 & 1RXSJ152912.9-101623 & $ 6709.926\pm 0.002 $ double-eclipse \\
0037737816 & MGAB-V203 & $12858.4\pm 0.4$ double-eclipse \\
0041742868 & SDSSJ153213.67+370104.8 & $ 28849\pm 2 $\\
0048014849 & CRTS\_SSS100621\_094840.0-345138 & $ 6836.1\pm 0.2 $\\
0048446609 & HS1857+5144 & $23011.202\pm 0.007$ eclipse \\
0048621773 & ASASSN-14jm & $ 14975\pm 18 $\\
0049001541 & CRTSJ105640-312212 & $ 6604.1\pm 0.3 $ double-eclipse\\
0049614493 & 2MASSJ06215073-2908421 & $ 12425.57\pm 0.06$\\
0049707471 & DDE79 & $ 13716.0\pm 0.6 $\\
0050721797 & ASASSN-13dc & $ 26778\pm 100$\\
0050772124 & V0495Cas & $ 6142.4\pm 0.8 $\\
0051122455 & 2MASSJ07155389+5816065 & $ 11695.2\pm 0.3 $\\
0052591659 & 2MASSJ05371478+1554172 & $ 17481.8\pm 0.2 $\\
0052949934 & Gaia19cfg & $ 20134\pm 2 $\\
0054477921 & MTPup & $ 11981.51\pm 0.08 $\\
0054527638 & NSV15272 & $ 752.17\pm 0.07 $\\
0055502089 & RXJ0636.3+6554 & $ 6154.1\pm 0.2 $ double eclipse\\
0058017207 & FBS0019+348 & $ 19922.1\pm 0.6 $\\
0058719633 & SDSSJ001204.50+020129.8 & $ 11816\pm 5 $ eclipse\\
0060952336 & ASASSN-15ay & $ 13370\pm 3 $\\
0061423366 & TXTri & $ 23104\pm 163 $\\
0064241826 & V1595Sgr &  $ 13237\pm 11 $\\
0067053526 & DKLac &  $ 10060.33\pm 0.07 $\\
0069465301 & V0923Cyg &  $ 22371.4\pm 0.4 $\\
0070700115 & ASASSN-18fp &  $ 41842\pm 49 $\\
0070834337 & MGAB-V236 &  $ 72580\pm 112 $\\
0071274998 & NSV13999 &  $ 13691.92\pm 0.11 $\\
0071448566 & CRTSJ102042.2-335002 &  $ 12727.8\pm 0.3 $\\
0071544757 & GSC07196-00489 &  $ 10896\pm 9 $\\
0071862439 & CRTSCSS120330J154450-115323 &  $ 12131\pm 15 $\\
0072842757 & ASASSN-19la  & $ 18973.6\pm 0.4 $ double-eclipse\\
0073435666 & CRTSSSS120320:101854-400644 &  $ 7110.79\pm 0.05 $\\
0073510290 & ASASSN-18os &  $ 170854\pm 106 $ double-eclipse\\
0073535720 & RXJ1404.4+1723  & $ 5690\pm 4 $\\
0074342209 & ASASSN-VJ194125.07+152255.4 & $ 5603.4042\pm 0.0010 $ eclipse\\
0075400109 & DDE44  & $ 79494\pm 582 $\\
0076327521 & Karachurin14  & $ 10844.12\pm 0.05 $\\
0077319633 & V0872Per &  $ 24399\pm 35 $\\
0079492961 & ASASSN-15hv &  $ 17150.4\pm 0.2 $\\
\hline
\end{tabular}
\caption{New Photometric Periods. Only a portion of this table is shown here to demonstrate its form and content. A machine-readable version of the full table is available in file kepler26.n 
%doi:10.5281/zenodo.21208066
(Kepler.new\_h for periods in hours). \label{newp}}
\end{table}

\end{document}